\documentclass{aa}  
\pdfoutput=1
\usepackage{graphicx,xcolor,natbib}
\usepackage{txfonts}
\usepackage[utf8]{inputenc}
\usepackage{xspace}
\usepackage{ulem} 
\usepackage{soul}
\usepackage{hyperref}
\usepackage{float}
\usepackage{longtable}
\usepackage{booktabs} 
\usepackage{lscape}
\usepackage{caption}
\usepackage[skip=5pt, belowskip=0pt]{caption}

\defcitealias{Bale2025}{Paper~I}

\begin{document} 

   \title{Chromospherically active stars: Lithium and CNO abundances in northern RS~CVn stars }

   \author{  B.~Bale\inst{1}
   \and
   G.~Tautvai\v{s}ien{\. e}\inst{1}  
     \and
          R.~Minkevi\v{c}i\={u}t\.{e}\inst{1}
         \and
          A.~Drazdauskas\inst{1}
          \and
          \v{S}.~Mikolaitis\inst{1}
          \and
          E.~Stonkut\.{e}\inst{1}
          \and
          M.~Ambrosch\inst{1}
          }

   \institute{Vilnius University, Faculty of Physics, Institute of Theoretical Physics and Astronomy, Saul\.etekio av. 3, 10257 Vilnius, Lithuania \\
   \email{barkha.bale@ff.vu.lt}
   }

   \date{Received ...; accepted  ...}

 
  \abstract
   {}
  {We carried out a detailed investigation of Lthium and CNO abundances, including carbon isotope ratios, in RS\,CVn stars to assess the role of magnetic activity in the mixing of stellar atmospheres. }    
{ We obtained high-resolution spectra at the Moletai Astronomical Observatory. Lithium abundances were determined by spectral synthesis of the  6707~{\AA} line and the CNO abundances using the ${\rm C}_2$ band heads at 5135 and 5635.5~{\AA}, CN bands at 6470--6490~{\AA} and 7980--8005~{\AA}, and the [O\,\textsc{i}] line at 6300~{\AA}. By fitting the $^{13}$CN band at 8004.7~{\AA}, we determined the carbon isotope ratios. }            
{We determined the main atmospheric parameters and investigated the chemical composition of 32 RS\,CVn stars. Lithium abundances were determined for 13 additional stars using archival spectra. We report that *iot\,Gem and HD\,179094 have carbon isotope ratios already affected by extra-mixing, even though they are in the evolutionary stage below the red giant branch luminosity bump. 
About half of the low-mass giants, for which the lithium abundance was determined, follow the first dredge-up predictions; however, other stars show reduced Lithium abundances, as predicted by thermohaline-induced mixing. The intermediate-mass stars show reduced Lithium abundances reduced, as predicted by rotation-induced mixing.}
{In low-mass, chromospherically active RS\,CVn stars, extra-mixing of lithium and carbon isotopes may begin earlier than in normal giants. 
The Li-rich RS CVn giant V*OP And has large C/N
and carbon isotope ratios and raises questions about the origin of its lithium enhancement. }


\titlerunning{Lithium and CNO abundances in northern RS~CVn stars}
\authorrunning{Bale et al.}

\maketitle

\section{Introduction}

Lithium occupies a unique niche in astrophysics owing to its intricate nucleosynthetic origins, its susceptibility to destruction during stellar evolution, and its role as an indicator of stellar interior conditions. In the early Universe, Big Bang nucleosynthesis (BBN) forged the lightest nuclei, synthesising deuterium, $^3$He, $^4$He, and $^7$Li. A widely accepted primordial Lthium abundance of $A({\rm Li})\simeq2.72$~dex was estimated by \cite{Cyburt2008} and later increased to $A({\rm Li})\simeq2.75$~dex by \citep{Pitrou2018}. However, observations of halo dwarfs ($-2.4 <{\rm [Fe/H]}<-1.5$, $5500<T_{\rm eff}<6250$~K) by \cite{Spite&Spite1982} yielded a smaller and constant $A({\rm Li})\simeq2.05$~dex, the so-called `Spite plateau'. \cite{Norris2023} comprehensively reviewed later developments in primordial lithium, the Spite plateau, and `lithium meltdown' investigations. Here, we highlight the discovery of a thin lithium plateau at $A({\rm Li})\simeq1.09\pm0.01$~dex among metal-poor, lower red giant branch (RGB) stars \citep{Mucciarelly2022}. Researchers also discovered a small fraction of Li-poor stars with $A$(Li) lower than $\sim0.7$~dex.
The lithium observed in the present-day Universe
is only in part the Lithium that was originally produced during
the Big Bang. Its abundance is modified by a number of
constructive and destructive processes that make Lithium one of the
elements with the most complex histories. Measuring the amount of lithium depletion requires knowledge of both the initial and current abundances.
The abundance of $A$(Li)\footnote{$
A(\mathrm{Li}) = \log\;\Bigl(\frac{X(\mathrm{Li})}{X(\mathrm{H})}\;\cdot\;\frac{A_{\mathrm H}}{A_{\mathrm Li}}\Bigr) + 12,
$} = 3.26~dex, currently found in the interstellar medium and Solar System meteorites, is considered the
reference limit for Population~I dwarf stars \citep{Asplund2009}, while the Lithium abundance of red giant stars is expected to decrease
to a value $A$(Li)$\sim1.5$~dex or less.

Lithium abundances gradually begin to decrease during the pre-main sequence, then decline further in the main sequence, during the first dredge-up (1DUP) at the base of the RGB, and then again at the RGB luminosity bump (see, e.g. the reviews by \citealt{Pinsonneault1997, Lyubimkov2016, Randich2021}). 

Lithium is fragile in stellar interiors, surviving only up to
$\sim2.5 \times 10^6$~K at typical envelope main-sequence densities. 
During the 1DUP process \citep{Iben1968}, the convective envelope deepens significantly, combining cooler lithium-containing material with hotter lithium-depleted layers, diluting the observable photospheric lithium abundance. 

Studies of stellar mixing have examined both rotational effects \citep{Palacios2003,Chaname2005} and effects of magnetic fields \citep{Busso2007,Nordhaus2008}. Thermohaline mixing, as detailed in \cite{Eggleton2006,Eggleton2008,Charbonnel&Zahn2007, Cantiello&langer2010, Charbonnel&Lagarde2010,Charbonnel2017}, adds further complexity by interacting with rotation. 

Moreover, observations show that lithium can also be produced by stars or acquired externally (\citealt{Smiljanic2018, Casey2019, Magrini-Lagarde2021, Martell2021, Soares2021, Sayeed2024} and references therein).
About 1.2--2.2\% of red giants have lithium levels that exceed theoretical estimates \citep{Gao2019, Casey2019, Charbonnel2020}. Proposed mechanisms include Lithium production via the Cameron–Fowler mechanism in asymptotic giant branch (AGB) stars \citep{Cameron&Fowler1971, Sackmann&Boothroyd1992}, nova outbursts \citep{Starrfield1978, Vigroux&Arnould1979, Izzo2015, Molaro2016, Rukeya2017}, and cosmic-ray spallation \citep{Reeves1970, Olive&Schramm1992}. Several studies have attributed lithium enrichment to a particular phase of stellar evolution \citep{Kumar2020, Mallick2023} or to alternative mechanisms such as rotation-driven mixing \citep{Denissenkov&Herwig2004}, magnetic buoyancy \citep{Busso2007,Nordhaus2008,Guandalini2009}, engulfment of planets \citep{Villaver&Livio2009,Adamow2012}, and interactions with substellar companions \citep{King1997, Israelian2004}.  Many other studies and reviews have explored these and other channels, but no single mechanism alone can explain the full diversity of observed lithium abundances.

Our study aimed to investigate how stellar magnetic activity influences the lithium content in the atmospheres of RS~Canum Venaticorum (RS\,CVn) stars. First characterised by \cite{Hall1976} and later refined by \cite{Fekel1986},  RS\,CVn systems comprise close binaries in which at least one cool component exhibits vigorous magnetic activity. This activity produces dark photospheric spots that cover significant fractions of the stellar surface and strong chromospheric emission lines \citep{Walter1980}. Consequently, RS\,CVn–type stars exhibit significantly greater activity than single stars of identical mass and age \citep{Cao2025}. These systems display intense multi‐wavelength activity, including coherent radio emission, X‐ray flares, and optical variability, often manifested as energetic flare events that reflect their dynamic atmospheres and complex magnetic fields (e.g. \citealt{Rucinski1993, White1978, Toet2021, Cao2023}). They also show a pronounced emission in Ca\,\textsc{ii} H\&K lines and the Balmer H$_\alpha$ line, a hallmark of chromospheric activity that earned them the designation of chromospherically active binaries.  
\cite{Popper1977} demonstrated that RS\,CVn systems are post-main-sequence objects that have just recently evolved out of the main sequence. Unlike main-sequence objects of spectral types later than about G8 to K0, where convective mixing largely depletes Li, the detection of the Li\,\textsc{i}\,6707\,\AA\ resonance doublet in RS\,CVn primaries (which is otherwise extremely rare in stars of types later than about G8 due to increasingly deep convective zones) signifies incomplete depletion caused by hindered mixing from rapid rotation or recent lithium production or enrichment \citep{Herbig1965,Soderblom1984,Randich1992}.

In a survey of southern RS\,CVn binaries and other chromospherically active stars, \citet{Pallavicini1992} measured the Li\,\textsc{i}\,6707\,\AA\ line and used curves of growth in a sample dominated by K-type stars. They found that many RS\,CVn stars exhibit high lithium abundances well above those of inactive field stars of the same spectral type. To verify this finding, \citet{Randich1993} performed a detailed spectrum synthesis of the RS\,CVn stars and confirmed a systematic lithium overabundance relative to field giants. They found a clear dependence of lithium on effective temperature, but only weak correlations with projected rotational velocity and chromospheric activity, and found no spectroscopic evidence for fresh lithium  production via spallation. \citet{Randich1994} used high‐resolution spectra of northern RS\,CVn stars to search for flare‐induced lithium signatures, but detected none. Their results showed that tidal locking and enhanced magnetic activity suppress deep convective mixing, inhibiting lithium depletion, and explaining the enhanced surface abundances without invoking ongoing lithium synthesis. They also suggested that more information can be obtained by comparing the lithium and carbon isotope $^{12}$C/$^{13}$C ratio of chromospherically active stars with normal inactive giants and subgiants. 

Notably, earlier investigations of RS\,CVn stars frequently lacked measurements of CNO abundances, carbon isotope ratios, and stellar mass determinations, which are critical for studying mixing in evolved stars. Our paper fills this gap by presenting these measurements.

We observed high-resolution spectra for a sample of 32 RS~CVn stars and investigated the abundances of lithium, carbon isotopes $^{12}$C and $^{13}$C, nitrogen, oxygen, and magnesium. In addition, we used the archival spectra from \citet{Bale2025} (hereafter Paper~I), in which we determined the atmospheric parameters, CNO abundances, and carbon isotope ratios, to derive the lithium abundances for 13 additional RS~CVn stars. 

The layout of this work is as follows. Sect. \ref{sample} describes how we selected our RS\,CVn targets, obtained high-resolution spectra, and derived stellar parameters, element and carbon isotope abundances, masses, and kinematics. Section~\ref{Resultsdis} outlines our empirical findings and their interpretation: an examination of extra-mixing in RS\,CVn stars, evaluation of chemical abundance trends, comparative analysis with theoretical models and observations of inactive giants, and a discussion of how magnetic activity shapes the evolution of RS\,CVn systems. Section~\ref{conclusion} summarises our key results and conclusions.

\section{Stellar sample and analysis }

We used the same instruments for the observations and the analysis method as in \citetalias{Bale2025}. Here, we present key details, including sample selection and our method for determining lithium abundances.

\label{sample}
\subsection{Sample selection}
\label{sample selection} 

We selected most of the sample from \cite{Strassmeier1994}, including the stars *iot\,Gem, *bet\,Gem, *g\,Gem, HD\,71028, V*\,FI\,Cnc, *10\,LMi, V*\,FG UMa, *tau\,Leo, *93\,Leo, *c\,Vir, V*\,IT\,Com, *7\,Boo, *eta\,Boo, V*\,FR Boo, HD\,141690, HD\,161832, HD\,179094, *f\,Aql, V*\,V1971\,Cyg, and *alf\,Sge. We selected the stars *33\,Psc, V*V1149\,Ori, *sig\,Gem, V*\,BM\,CVn, V*\,HK\,Boo, V*\,V835\,Her, V*\,V2075\,Cyg, V*\,IM\,Peg, *lam\,And, and *eps\,Umi from \cite{Eker2008}. The star V*OP\,And was included from \cite{Strassmeier1990}, and the star HD\,6497 was taken from \cite{Massarotti2008}. We chose these stars based on a variety of criteria, including brightness, positional coordinates, and the allocated observation time. In the case of spectroscopic binaries, only single-lined stars were selected.

\subsection{Observations}
\label{Observation} 

The stellar observations were carried out from 2021 to 2024 using the 1.65~m Ritchey–Chrétien telescope at Vilnius University’s Moletai Astronomical Observatory. We acquired high-resolution spectra with the Vilnius University Echelle Spectrograph (VUES) \citep{Jurgenson2016}, covering a spectral range from 4000 to 8800~(\AA\}. We used spectral resolutions of $R\sim 36\,000$ and $R\sim 68\,000$. The stars were exposed for 1200 to 3600~s depending on brightness, resulting in signal-to-noise ratios (S/N) ranging from 57 to 292, with a median S/N of approximately 136. Data reductions were performed using the automated pipeline developed specifically for the VUES spectrograph by \cite{Jurgenson2016}.

\begin{figure*}
    \centering
        \includegraphics[width=1.0\textwidth]{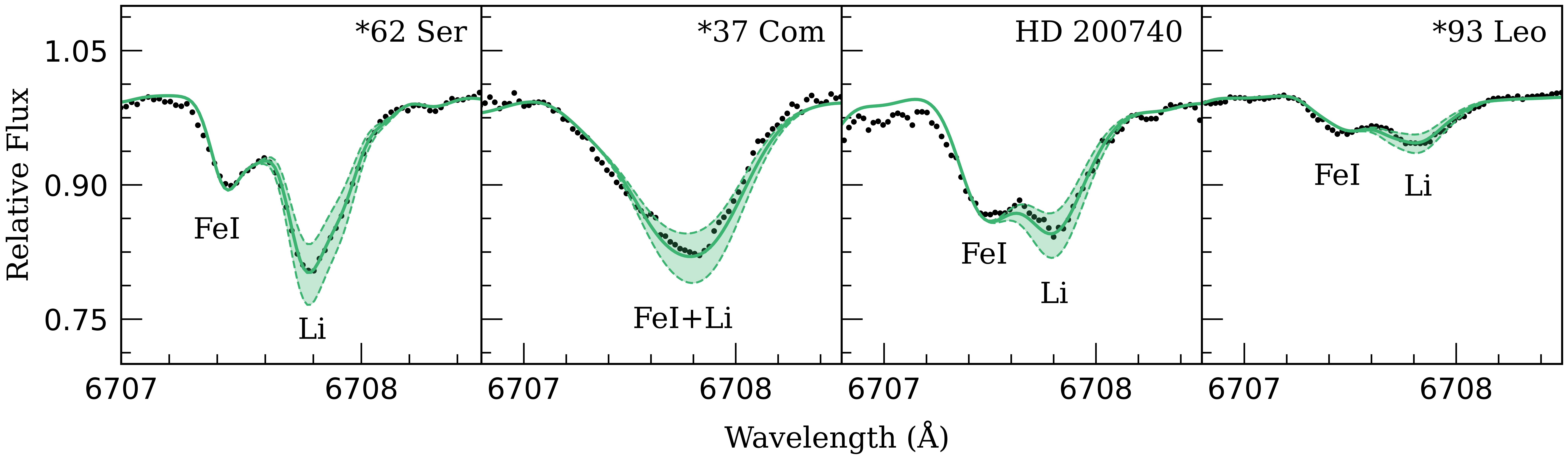}
        \includegraphics[width=1.0\textwidth]{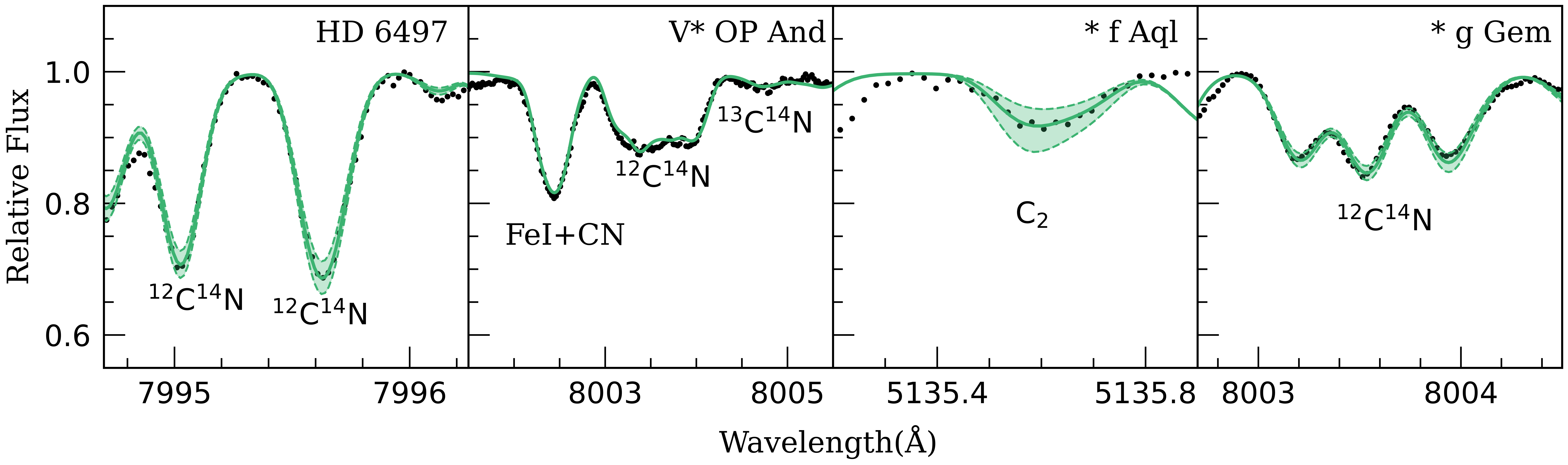}
    \caption{Examples of synthetic spectrum fits to the Li\,\textsc{i} line, C$_2$ band heads at 5135\,\AA, and the CN bands at 7995\,\AA{} and 8003\,\AA. Black dots represent the observed spectra, while solid green lines depict the best-fit synthetic spectra. The shaded region around the fit indicates the $\pm0.10$~dex abundance range.}
    \label{4o}
\end{figure*}

We obtained a total of 41 observations in two spectroscopic modes for 32 objects: 17 stars were observed at $R\sim 36\,000$, six at $R\sim 68\,000$, and the remaining nine in both configurations. Details of the spectral resolution used are presented along with the results.  
As the lithium abundance was not determined for stars investigated in \citetalias{Bale2025}, we used the archival spectra to derive lithium abundances,  successfully determining them for 13 stars. 

\subsection{Atmospheric parameters}
\label{Atmospheric parameters} 

To determine stellar atmospheric parameters, we considered the effective temperature ($T_{\rm eff}$), surface gravity ($\log g$), microturbulent velocity ($v_{\rm t}$), and overall metallicity. We used the DAOSPEC code \citep{Stetson&Pancino2008} to extract equivalent widths for Fe\,\textsc{i} and Fe\,\textsc{ii} lines. The MOOG code \citep{Sneden1973} was used for the iterative adjustment of $T_{\rm eff}$, $\log g$, $v_{\rm t}$, and [Fe/H] until there was no slope for [El/H] versus excitation potential or [El/H] versus equivalent width for both Fe\,\textsc{i} and Fe\,\textsc{ii} lines. On average, 58 Fe\,\textsc{i} and six Fe\,\textsc{ii} lines were used for the analysis. Atomic lines were selected from the Gaia-ESO line list of \cite{Heiter2015}. 
The model atmospheres were taken from the Model Atmospheres with a Radiative and Convective Scheme (MARCS) stellar model atmosphere and flux library \citep{Gustafsson2008}, and we adopted the solar abundances of \cite{Grevesse2007}.

\subsection{Elemental abundances and carbon isotope ratios}
\label{elemental abu& cc ratio para}

\begin{figure}
    \includegraphics[width=\columnwidth]{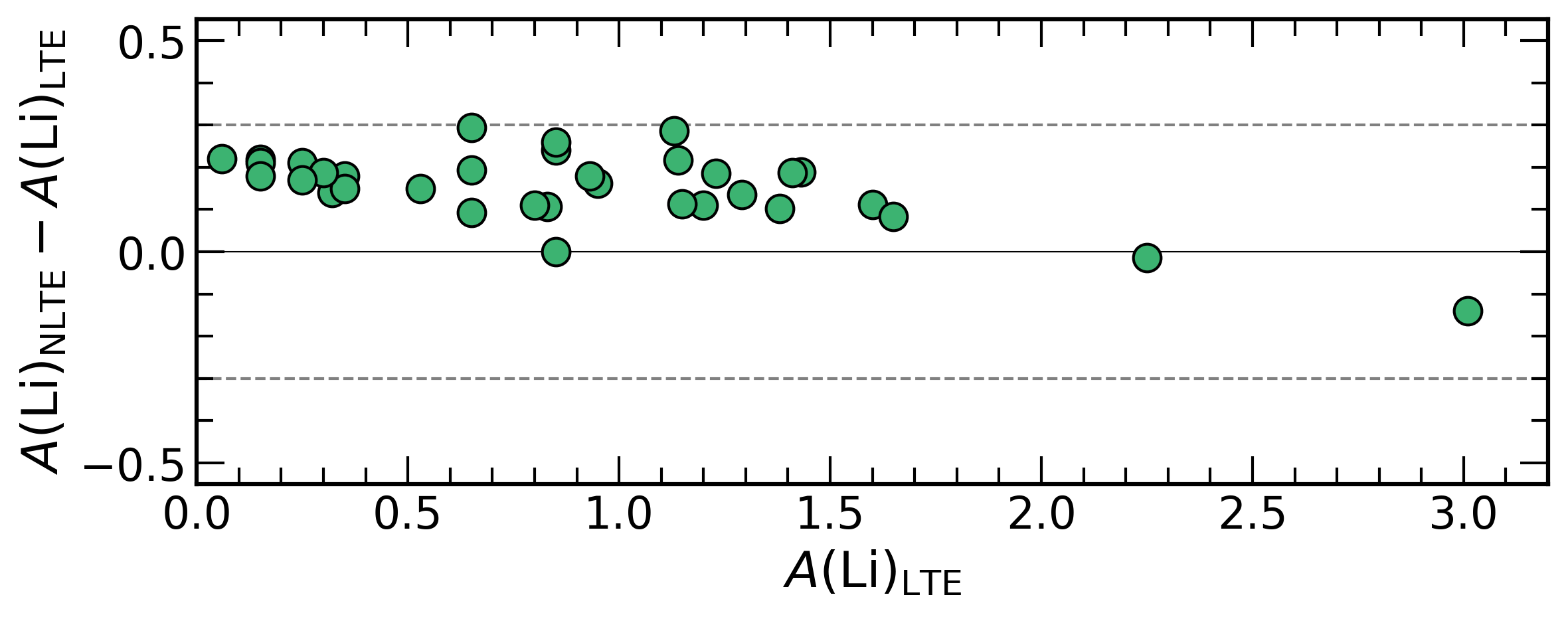}
    \caption{Comparisof NLTE vs LTE lithium abundances. }
    \label{ltevs}
\end{figure}

We derived the abundances of Li, C, N, O, and Mg by spectral synthesis with the \texttt{TURBOSPECTRUM} code \citep{Plez2012}. Figure~\ref{4o} illustrates examples of fits of the spectral syntheses to several spectral features. We determined lithium abundances from the 6708\,\AA\ line and applied non–local thermodynamic equilibrium (NLTE) corrections using the INSPECT database \citep{Lind2009}\footnote{\url{http:https://www.inspect-stars.com/}}. Fig.~\ref{ltevs} shows a comparison of NLTE and local thermodynamic equilibrium (LTE) lithium abundances. For stars with lithium abundances below 2~dex, the corrections are positive and reach about 0.3~dex, while for stars with higher lithium abundances, the corrections become negative.

We measured oxygen abundances from the forbidden [O\,\textsc{i}] line at 6300.3\,\AA. and we obtained carbon abundances from the C$_2$ molecular bands at 5135\,\AA\ and 5635\,\AA. Nitrogen abundances were determined using $^{12}$C$^{14}$N lines in the 6470–6490\,\AA\ and 7980–8005\,\AA\ regions. We used the 7990–8010\,\AA\ interval, containing $^{13}$C$^{14}$N features, to derive the carbon isotope ratio.
We performed several iterative steps to calculate the abundances of carbon and oxygen, which are linked by molecular equilibrium.   We then fixed the final C and O values to derive N abundances and the $^{12}$C/$^{13}$C ratio.  A comprehensive description of the methodology and procedures is available in the Gaia-ESO paper by \cite{Tautvaisiene2015}. Although the NLTE effects on the C$_2$ Swan bands remain poorly characterised, they appear negligible, since abundances from the forbidden [C\,\textsc{i}] line agree with those of C$_2$ \citep{Gustafsson1999}. The [O\,\textsc{i}] 6300.3\,\AA\ line is well modelled under LTE \citep{Asplund2005}.

 In V* HK Boo,  HD 141690, V* V835 Her, V* V1971 Cyg, and *eps UMi, the [O\,\textsc{i}] spectral lines were contaminated by telluric interference, which affected the direct determination of the oxygen abundance. We therefore accepted magnesium abundances determined from Mg\,\textsc{i} lines at 5528.41\,\AA, 5711.07\,\AA, 6318.71\,\AA, and 6319.24\,\AA  as an alternative indicator of the oxygen abundance in these stars.

\subsection{Stellar masses}
\label{Stellar masses} 
We estimated stellar masses with the UniDAM tool \citep{Mints&Hekker2017,Mints&Hekker2018}, which combines Bayesian inference with PAdova and TRieste Stellar Evolution Code (PARSEC) isochrones \citep{Bressan2012}. Our input included spectroscopic atmospheric parameters and near-IR photometry ($J$, $H$, and $K$ colours from 2MASS, and $W1$ and $W2$ colours from AllWISE). When assigning the stellar mass values, we took into account the quality flags from Sect.~6.1. of \citet{Mints&Hekker2017,Mints&Hekker2018}.

\subsection{Kinematics properties}
\label{Kinematics} 
\begin{figure}
    \includegraphics[width=\columnwidth]{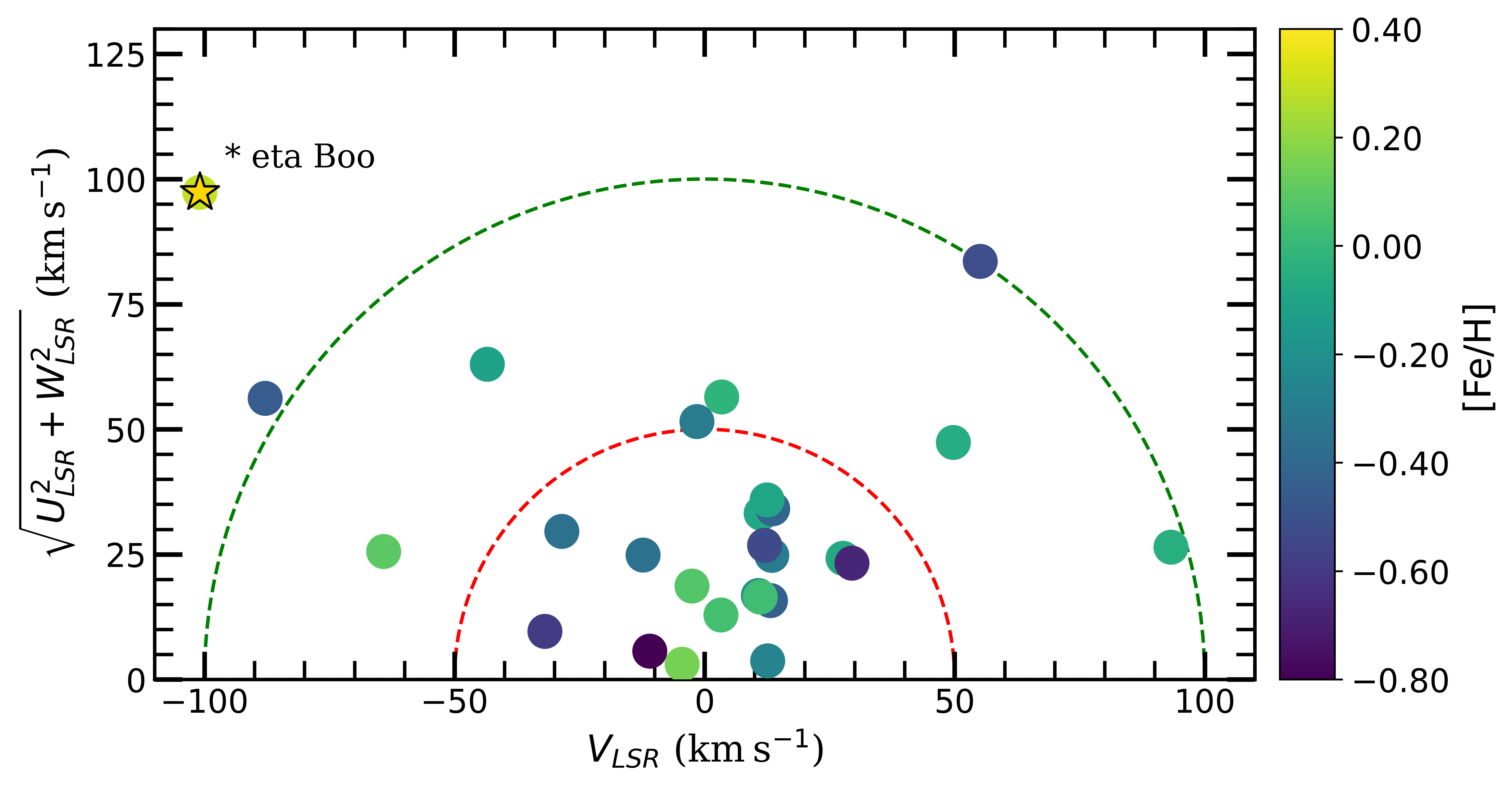}
    \caption{Sample stars colour-coded by metallicity on a Toomre diagram. Dashed lines indicate constant total space velocity. The star *eta\,Boo, attributed to the thick disc, is marked with a star symbol.}
    \label{uvw}
\end{figure}

We used the Python package \texttt{galpy}\footnote{\url{http://github.com/jobovy/galpy}} \citep{Bovy2015} to compute Galactic velocities $(U, V, W)$ based on distances from \citet{Bailer-Jones2021}, proper motions, coordinates, and radial velocities from \textit{Gaia} DR3 \citep{GaiaCollaboration2021}. A Toomre diagram (Fig.~\ref{uvw}) separates thin‐disc ($|V_{\rm tot}|<50\ \mathrm{km\,s^{-1}}$) from thick‐disc ($|V_{\rm tot}|>50\ \mathrm{km\,s^{-1}}$) stars following \citet{Yoshii1982}, \citet{Gilmore&Reid1983}, and \citet{Recio-Blanco2014}. By combining kinematic parameters and chemical abundances, we classify only *eta\,Boo as a thick‐disc member, while all other targets reside in the thin disc.

\subsection{Estimation of uncertainties }
\label{uncertainties} 

\begin{table}
\centering
\caption{Effect of uncertainties in atmospheric parameters on the derived chemical abundances for the star HD 179094.}
\label{tab:errtab}
\begin{tabular}{lccccccl}
\hline
\hline
El.     & $\Delta T_{\rm eff}$  & $\Delta {\rm log}\,g$  & $\Delta$[Fe/H]  & $\Delta v_{\rm t}$ \\
        &  $\pm$55 K  & $\pm$0.16 dex    & $\pm$0.09dex     &$\pm0.21~{\rm km\, s}^{-1}$ \\
\hline

\hline
Li    & $\pm$0.07   & $\pm$0.02  & $\mp$0.02  & $\pm$0.00 \\ 
C(C\textsubscript{2})     & $\pm$0.01   & $\mp$0.02    & $\mp$0.02  & $\mp$0.01 \\
N(CN)  & $\pm$0.02   & $\mp$0.04  & $\mp$0.03  & $\pm$0.01  \\
O([O\,\textsc{i}])     & $\pm$0.01   & $\pm$0.07 & $\mp$0.01  & $\pm$0.01 \\ 
Mg\,\textsc{i}   & $\pm$0.04   & $\mp$0.06  & $\mp$0.01  & $\mp$0.06 \\ 
$^{12}$C/$^{13}$C & $\mp$0.30  &$\mp$1  &  $\mp$0.30 &  $\pm$0.00 \\
\hline
\end{tabular}
\end{table}

In the present study, we systematically identified and quantified sources of uncertainty at every stage of our analysis, distinguishing between systematic and random components. Systematic errors were mainly due to uncertainties in the atomic data. We reduced these uncertainties by comparing our results directly with the Sun. Random uncertainties mainly arose from the placement of the local continuum and the fitting of each spectral line. To ensure robust atmospheric parameters, we chose an optimal set of Fe\,\textsc{i} and Fe\,\textsc{ii} lines that avoid blends, telluric lines, and challenging continuum zones.  The mean values of the uncertainties in the full stellar sample are \( \sigma_{T_{\text{eff}}} = 55 \, \text{K} \),  \( \sigma_{\log g} = 0.16 \,  \text{dex} \),  \( \sigma_{[\text{Fe/H}]} = 0.09 \,  \text{dex} \), and  \( \sigma_{v_t} = 0.21 \,  \text{km s}^{-1} \). The precision of elemental abundances is influenced by uncertainties in atmospheric parameters. To quantify these effects, we varied each atmospheric parameter in isolation, keeping the rest constant, and determined the resulting abundance deviations. The uncertainties derived for HD\,179094 are presented in Table \ref{tab:errtab}. We estimated random uncertainties from the line-to-line scatter in the abundances of nitrogen and magnesium. We evaluated the probable uncertainties for oxygen, the carbon isotope ratio, and, in the case of carbon, when only one line was measured, from continuum placement variations according to the S/N. 

\begin{figure*}
    \centering
    \includegraphics[width=0.34\textwidth]{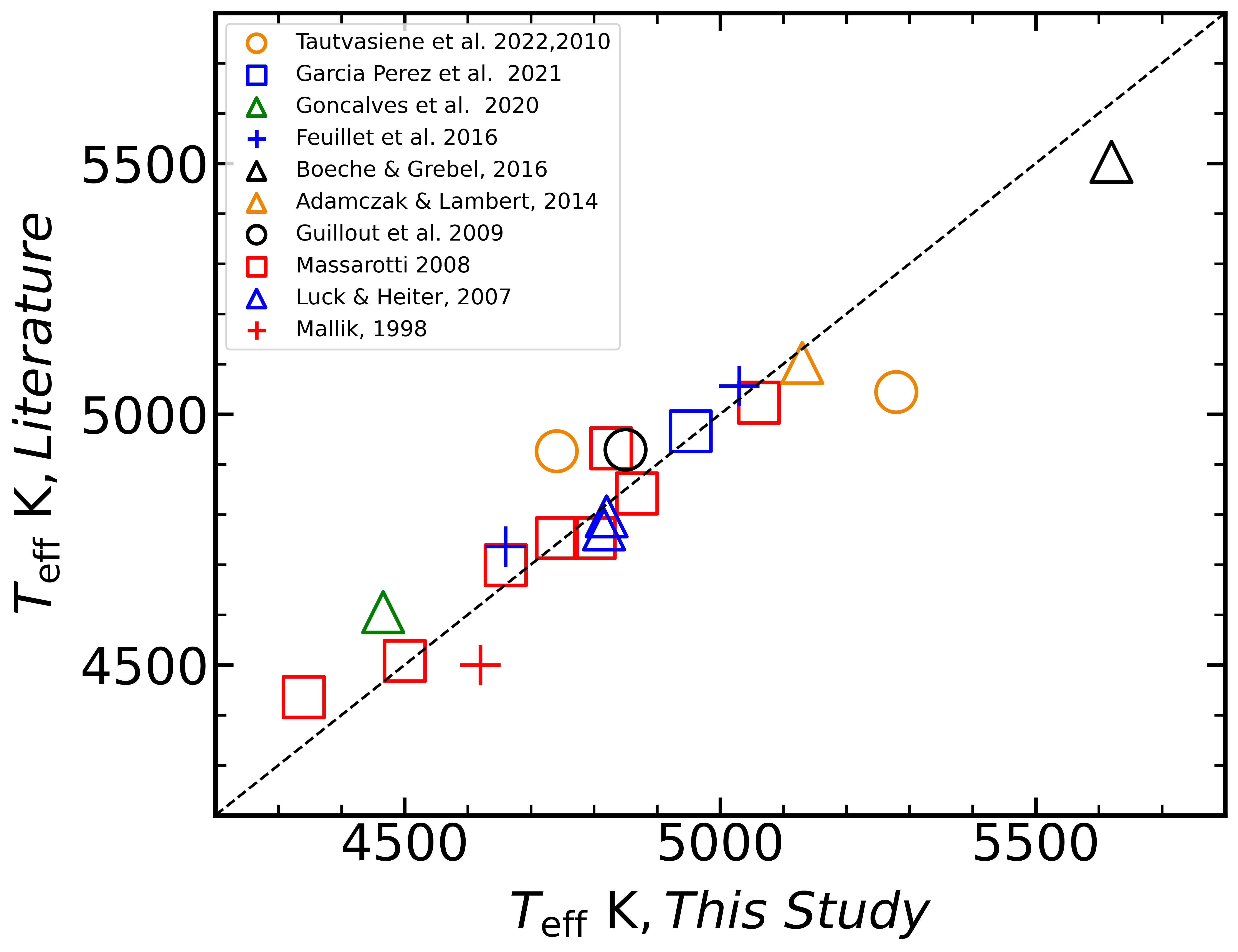}
    \includegraphics[width=0.31\textwidth]{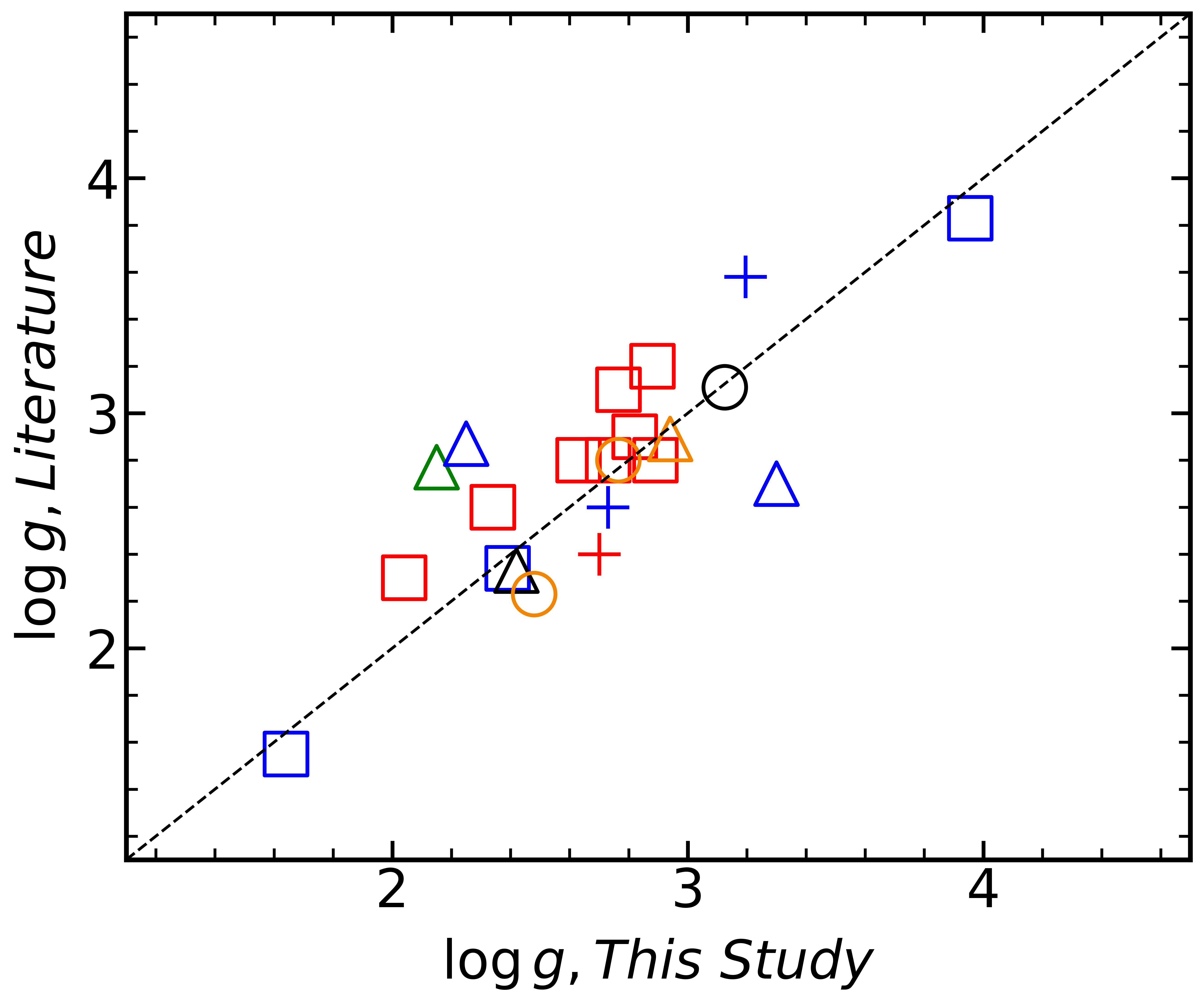}
    \includegraphics[width=0.34\textwidth]{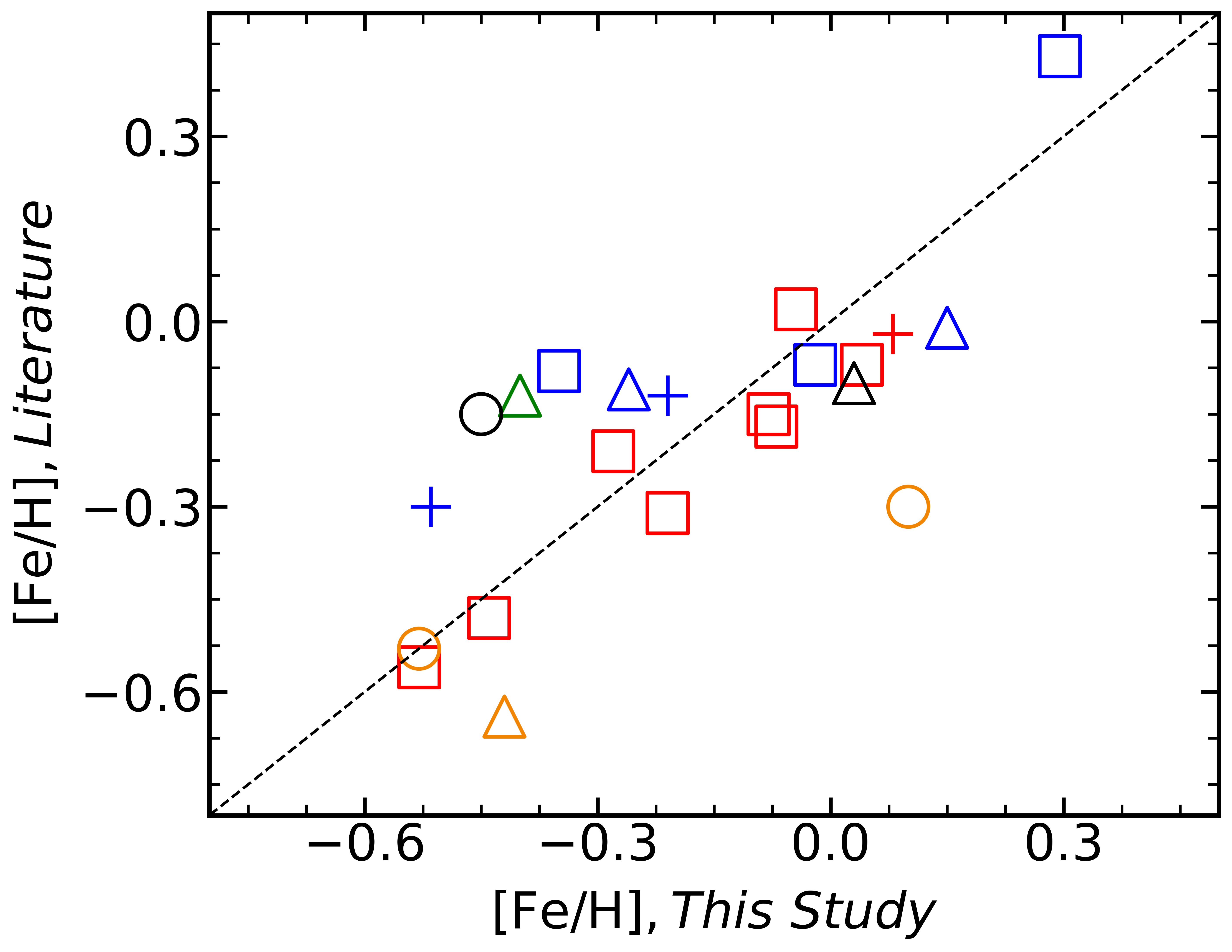}
    \caption{Atmospheric parameters derived in this study ($T_{\rm eff}$, left; $\log g$, middle; and [Fe/H], right) compared with values from studies listed in Sect.~\ref{resultstellar}.}
    \label{comppp}
\end{figure*}

Our analysis of molecular equilibrium effects on C, N, and O shows that  $\Delta$[O/H] = 0.10 drives a change of  $\Delta$[C/H] = 0.05, and  $\Delta$[N/H] = 0.02;  $\Delta$[C/H] = 0.10 induces a change of  $\Delta$[N/H] = 0.07, and  $\Delta$[O/H] = 0.04; while  $\Delta$[N/H] = 0.10 leads to  $\Delta$[C/H] = 0.03, and  $\Delta$[O/H] = 0.01.

We used the UniDAM tool (\citealt{Mints&Hekker2017, Mints&Hekker2018}) to compute uncertainties in stellar mass and age. The uncertainty values are provided together with the results in the machine-readable table~\ref{table:Results}.

\section{Results and discussion}
\label{Resultsdis}

\subsection{Stellar parameters}
\label{resultstellar} 
We conducted a systematic investigation of stellar objects spanning a range of post–main‐sequence evolutionary stages. 

In the machine-readable tables~\ref{table:Results} and \ref{table:Results2}, we provide all parameters determined in this study. 
The main atmospheric parameters of stars determined in this work are also provided in Table~\ref{table:atmospheric_parameters} for convenience.  
In all plots, we present the averaged values of parameters determined from spectra observed in the two spectral resolutions ($R=36\,000$ and $R=68\,000$), where available.

\begin{figure}
    \includegraphics[width=\columnwidth]{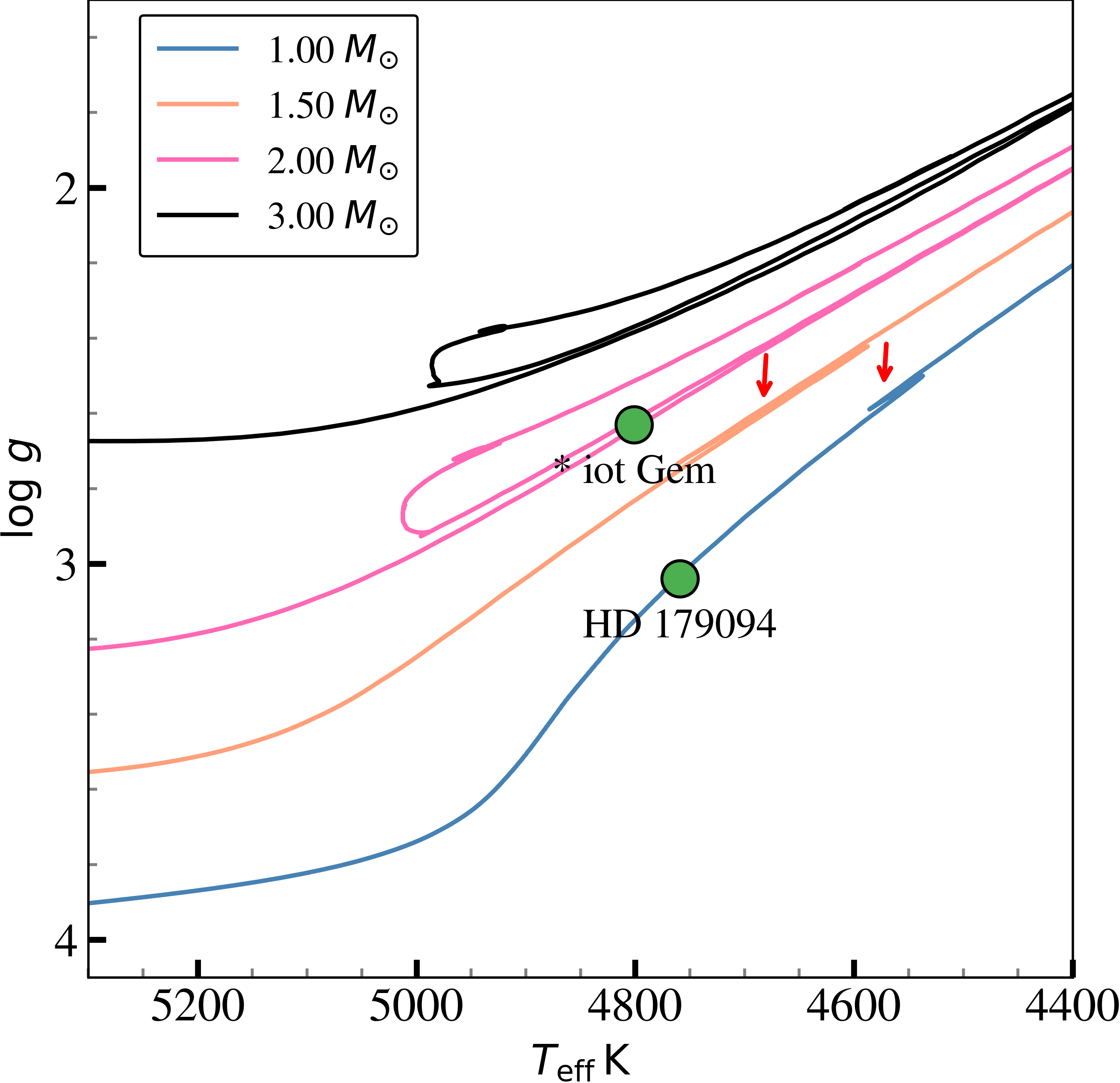}
    \caption{Log\,$g$ versus $T_{\rm eff}$ diagram. Locations of *iot\,Gem (1.37~$M_\odot$) and HD\,179094 (1.03 $M_\odot$) below the RGB luminosity bump are denoted by red arrows, on the PARSEC evolutionary tracks for solar metallicity from \cite{Bressan2012}.}
    \label{sett}
\end{figure}

The effective temperatures ($T_{\rm eff}$) of these stars lie between 4120 and 6220~K, the surface gravities (log $g$) range from 1.4 to 4.3~dex, and the metallicity ([Fe/H]) values span from  $-0.85$ to 0.34~dex, with an average of  $-0.26$~dex. 
Stellar ages cover a wide interval from 380~Myr to 10.5~Gyr; consequently the stellar masses range from 3.5~$M_\odot$ down to 0.9~$M_\odot$. 
Taken together, these quantities demonstrate that the RS\,CVn stars in our sample form a heterogeneous group of giants at different stages of their post‐main‐sequence evolution.

We compared our atmospheric parameters ($T_{\rm eff}$, log\,$g$, and [Fe/H]) with existing data from the literature (Fig.~\ref{comppp}). The comparison is made with spectroscopic studies by \citep{Tautvaisiene2022,GarciaPerez2021,Goncalves2020, Feuillet2016, Boeche&Grebel2016,Adamczak2014, Tautvaisiene2010a, Guillout2009,Massarotti2008, Luck&Heiter2007,Mallik1998}. The agreement in effective temperatures and surface gravities is good, whereas the scatter in metallicities is larger.

The lithium abundances, carbon isotope $^{12}$C/$^{13}$C ratios, and C/N ratios determined for 32 stars are presented in Table~\ref{tab:cccn1}. In Table~\ref{tab:cccn2}, the lithium abundances are presented for 13 stars investigated using archival spectra. For these stars, the $^{12}$C/$^{13}$C and C/N ratios determined in \citepalias{Bale2025} are also presented for convenience. In addition, we present stellar masses and evolutionary stages. We determined the evolutionary stages by inspecting the positions of stars in the log\,$g$ versus $T_{\rm eff}$ diagram,
using the PARSEC evolutionary tracks (\citep{Bressan2012}), and taking into account their carbon and nitrogen abundances. For stars with masses lower than 2.2\,$M_{\odot}$, the RGB luminosity bump, where the extra-mixing begins, is visible in the stellar evolutionary sequences as a zigzag. Stars with larger masses do not experience extra-mixing during the first ascent on the RGB, and the so-called second dredge-up only begins on
the asymptotic giant branch, unless rotation-induced mixing occurs \citealt{Charbonnel&Lagarde2010}.  Fig.~\ref{sett} shows the positions of two stars. We discuss these stars in Subsect.~\ref{CNOresult}.

In our sample of 32 stars, three are subgiants (we mark them as SG), 18 stars are below the RGB luminosity bump (we mark them as BB) and eight above (marked as AB). Two stars are attributed to the core He-burning red clump (marked as RC). 

\subsection{CNO abundances and carbon isotope ratios }
\label{CNOresult} 

The determination of the surface abundances of carbon and nitrogen in RS\,CVn stars offers a direct probe of how efficiently material is mixed within their convective envelopes under the combined influence of rotation and tidal forces.  In canonical stellar models, the 1DUP brings the processed material from the CN-cycle to the envelope, where $^{12}\mathrm{C}$ is depleted, whereas $^{13}\mathrm{C}$ and $^{14}\mathrm{N}$ become enhanced \citep{Iben1965, Charbonnel&Zahn2007}. Surface compositions undergo additional changes when extra-mixing begins at the RGB luminosity bump in low‐mass stars. The extent of the resulting change in $^{12}$C/$^{13}$C and C/N ratios depends on the evolutionary status of a star, its mass, and its metallicity \citep{Gilroy1989,Luck1994,Tautvaisiene2005,Tautvaisiene2010a,Smiljanic2009, Mikolaitis2010, Mikolaitis2011a, Mikolaitis2011b}.

\begin{figure}
    \includegraphics[width=\columnwidth]{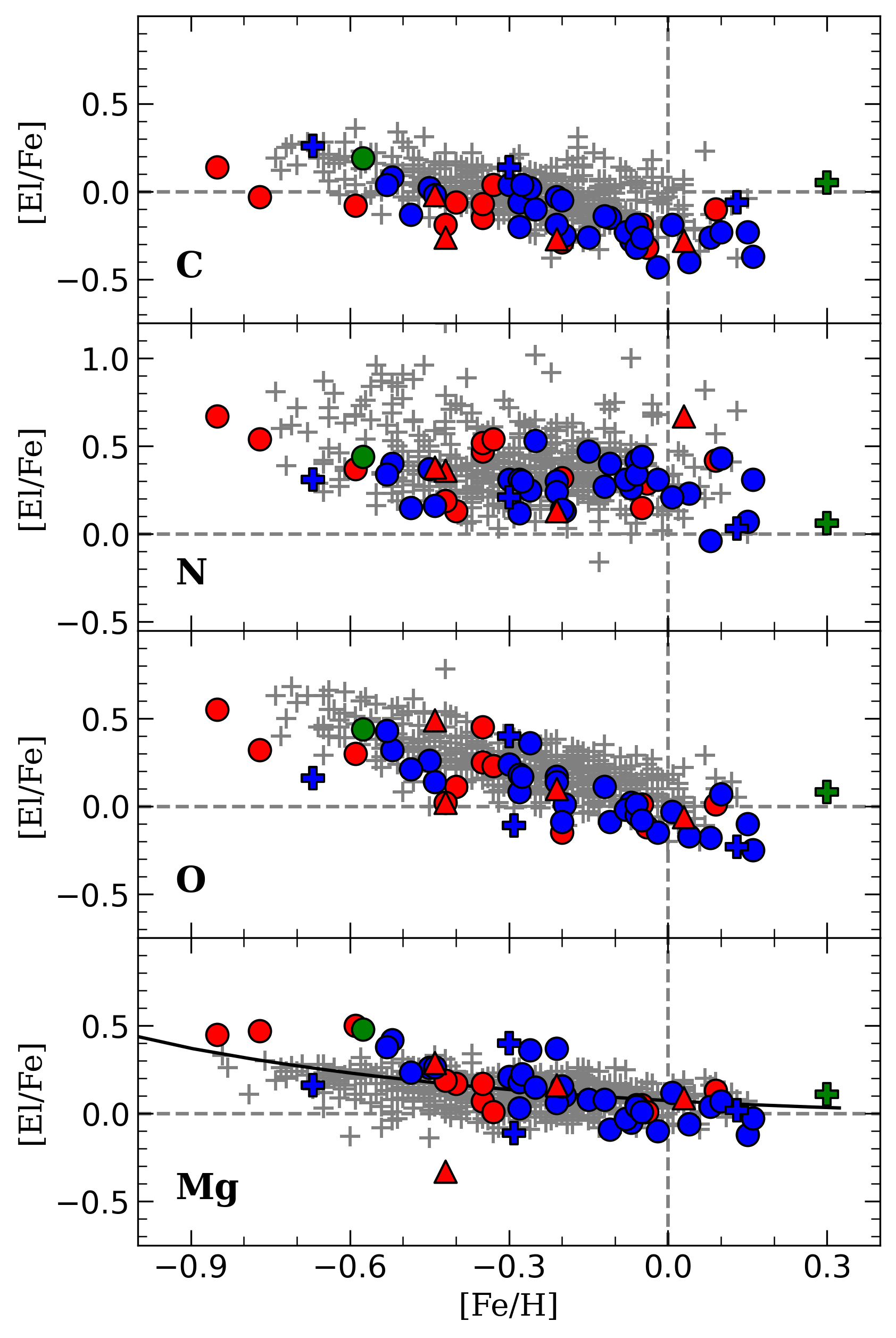}
    \caption{Element-to-iron abundance ratios from this work and \citetalias{Bale2025} as a function of [Fe/H]. Circles represent red giant branch (RGB) stars, plus signs indicate subgiant stars, and the triangles denote red clump stars. Blue symbols correspond to stars at the evolutionary stage below the RGB luminosity bump, while red symbols indicate stars above the bump. Green symbols mark stars of the thick disc. The solid line represents the Galactic chemical evolution model for the thin disc by \cite{Pagel1995}. Grey plus signs represent [El/Fe] abundances of thin-disc RGB stars from \cite{Tautvaisiene2022} for comparison.  } 
    \label{cnomg}
\end{figure}

\begin{figure}
    \includegraphics[width=\columnwidth]{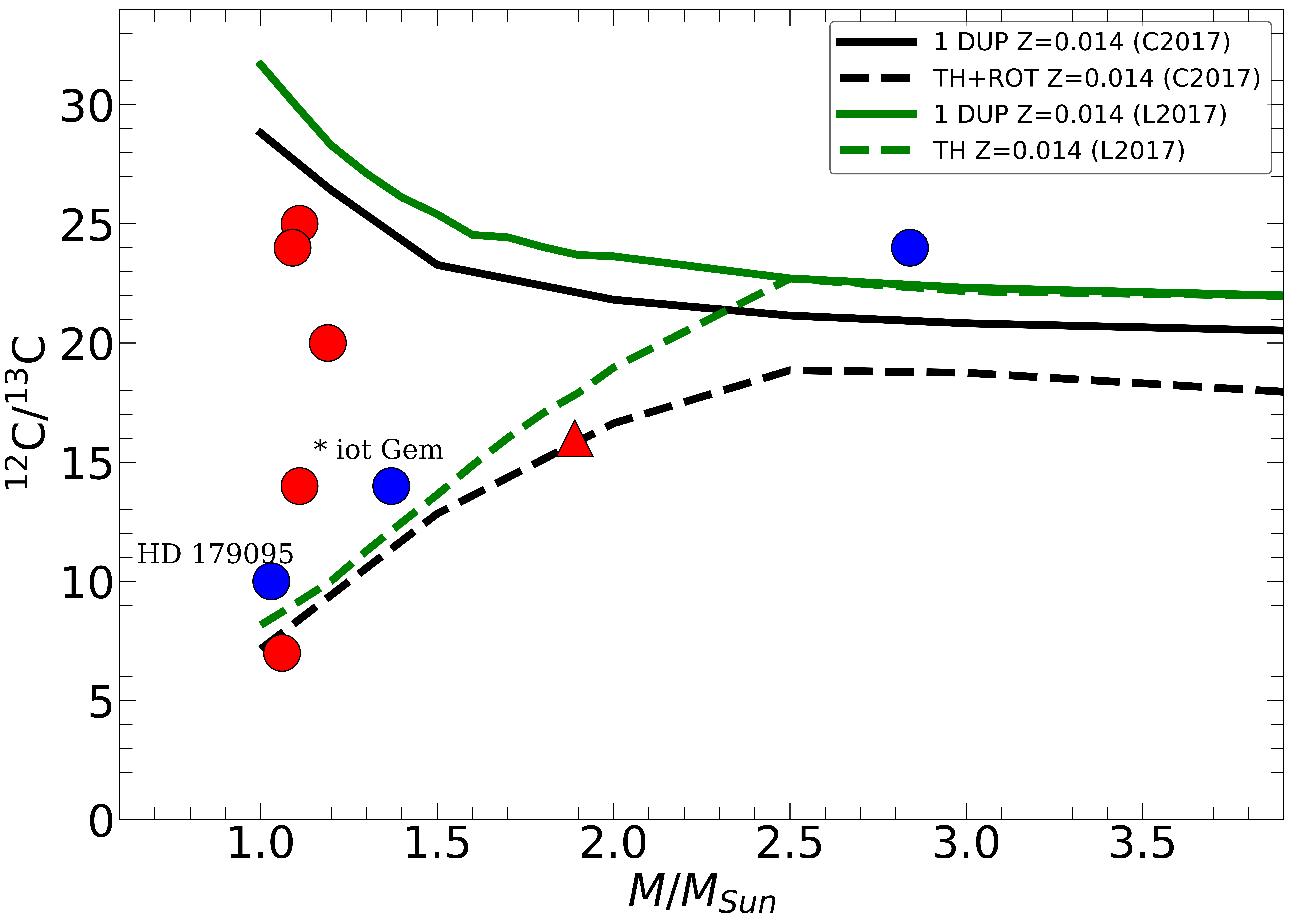}
        \includegraphics[width=\columnwidth]{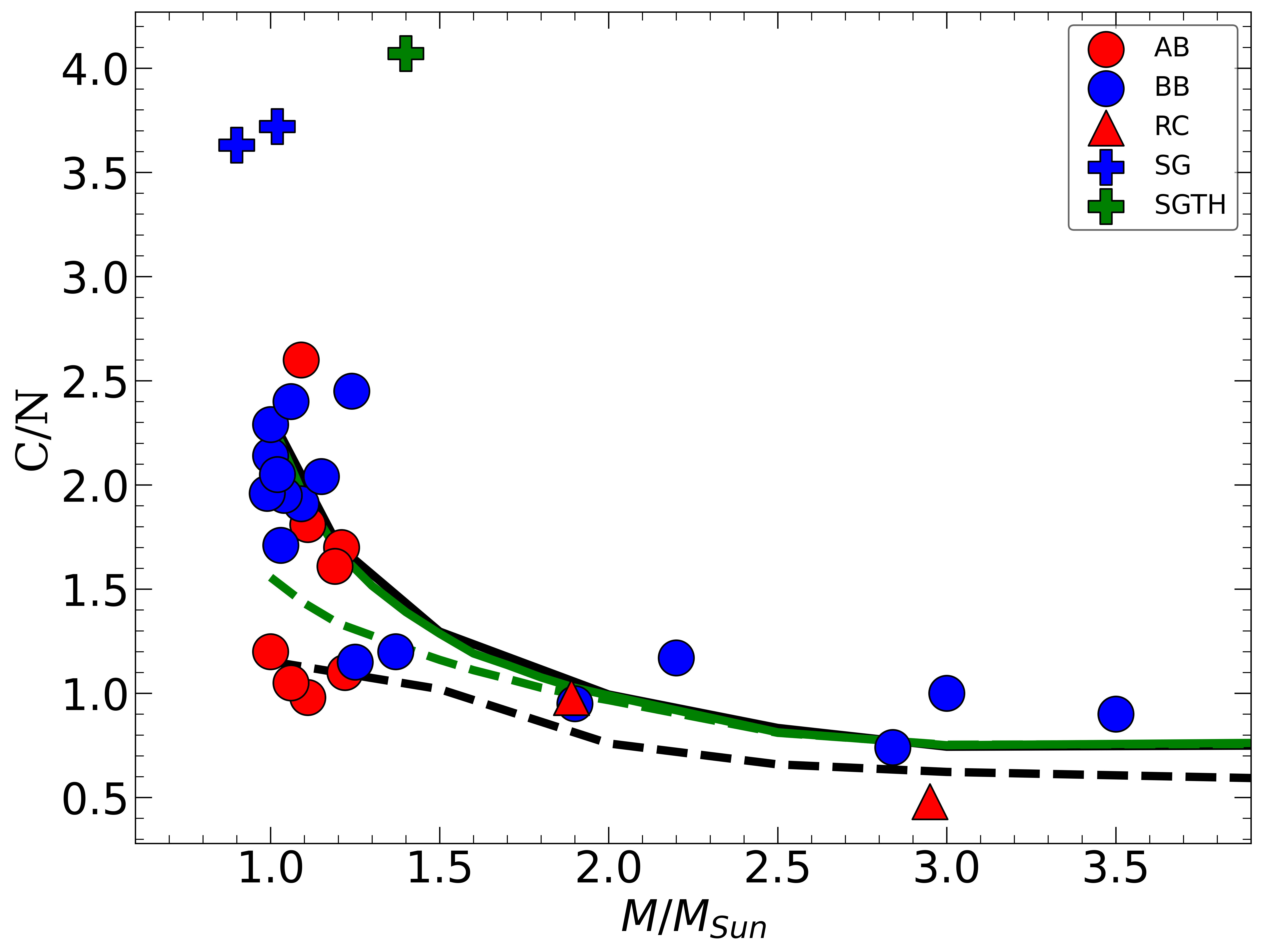}
    \caption {Comparison of \textsuperscript{12}C/\textsuperscript{13}C and C/N ratios with theoretical models. Solid green and black lines correspond to first dredge-up (1\,DUP) models from \cite{Lagarde2012} and \cite{Charbonnel2017}, respectively. The green dashed line represents the minimal values in the model with thermohaline-induced extra-mixing (TH) from \cite{Lagarde2017}, while the dashed black line shows minimal values in the model with thermohaline- and rotation-induced extra-mixing (TH+ROT) from \cite{Charbonnel2017}. Symbols are the same as in Fig.~\ref{cnomg}.}
    \label{CC}
\end{figure}

Figure~\ref{cnomg} presents the element‑to‑iron abundance ratios [El/Fe] as a function of [Fe/H], along with the results of \citetalias{Bale2025} and those for RGB thin-disc stars from \cite{Tautvaisiene2022}, which were determined using spectra from the same telescope and analysis method, for comparison. For magnesium, the predicted Galactic evolution trend from \cite{Pagel1995} is also shown. The investigated RS\,CVn stars at various evolutionary stages are plotted using different symbols. 
The giants below the RGB luminosity bump are shown as filled blue circles, the giants above it as filled red circles, the red clump stars as red triangles, and the thin-disc subgiants as blue crosses. The thick-disc subgiant star *eta\,Boo, attributed to this population according to its location in the Toomre diagram (Fig.~\ref{uvw}), is shown as the green plus sign. 

The chromospherically active RS\,CVn stars examined in this study do not exhibit systematic deviations in oxygen and magnesium abundances compared to inactive RGB stars or to the Galactic model \citep{Pagel1995}. The carbon abundances appear slightly more depleted, but the enhancement of the nitrogen abundance is not obvious across the sampled metallicity range. The thin-disc subgiants have slightly higher C and lower N abundances, as expected. The thick-disc subgiant shows slightly larger abundances of C and $\alpha$-elements, as expected.  

\begin{table}[ht]
  \centering
  \caption{Mass, $^{12}$C/$^{13}$C, C/N, NLTE lithium abundances, and evolutionary stage.}
  \label{tab:cccn1}
  \begin{tabular}{@{}lccclc@{}}
    \toprule
    Name                 & Mass      & $^{12}$C/$^{13}$C & C/N   & $A(\mathrm{Li})$ & Evol.        \\
                  & ($M_\odot$)&                   &       &                    NLTE       &   stage          \\
    \midrule
    *33 Psc  & 1.09	& --	& 1.91	& 0.28$^{({\ast})}$	& BB	 \\
HD 6497  & 1.11	& 24	& 1.58	& --	& AB 	\\ 
HD 6497$^{({\dagger})}$  & 1.11	& 26	& 2.04	& --	& AB	   \\ 
V* OP And$^{({\dagger})}$ & 1.09	& 24	& 2.60	& 2.87	& AB  \\ 
V* V1149 Ori  & 1.21	& --	& 1.70	& 1.09	& AB	  \\
*iot Gem & 1.37	& 14	& 1.20	& 0.36	& BB	  \\ 
*sig Gem$^{({\dagger})}$ & 1.24	& --	& 2.45	& 1.11	& BB	  \\ 
*bet Gem  & 1.90	& --	& 0.95	& 0.84  	& BB	\\ 
*g Gem & 1.11	& 14	& 0.98	& --	& AB	  \\ 
HD 71028 & 1.22	& --	& 1.10	& --	& AB	 \\ 
V* FI Cnc & 1.25	& --	& 1.15	& 1.71 	& BB\\ 
*10 LMi & 2.20	& --	& 1.17	& --	& BB	  \\
V* FG UMa & 1.00	& --	& 1.20	& 1.11	& AB	  \\ 
*tau Leo  & 2.84	& 24	& 0.74	& 0.49$^{({\ast})}$	& BB	 \\ 
*93 Leo$^{({\dagger})}$ & 1.89	& 16	& 0.98	& 1.31	& RC   \\
*c Vir   & 1.89	& 16	& 0.98	& --	& RC   \\ 
V* BM CVn  & 1.15	& --	& 2.04	& --	& BB \\ 
V* IT Com  & 1.00	& --	& 2.14	& 0.33$^{({\ast})}$	& BB	  \\ 
*7 Boo & 3.00	& --	& 0.93	& 0.96$^{({\ast})}$	& BB	\\ 
*7 Boo$^{({\dagger})}$ & 3.00	& --	& 1.07	& 0.87$^{({\ast})}$	& BB	  \\ 
*eta Boo$^{({\dagger})}$  & 1.40	& --	& 4.07	& --	& SG$^{({\ddagger})}$   \\ 
V* FR Boo & 1.06	& 7 	& 1.05	& --	& AB	 \\ 
V* HK Boo$^{({\dagger})}$ & 0.90   & --	& 3.63	& 1.73	& SG\\ 
HD 141690 & 1.15	& --	& --	& 2.24	& SG	  \\
HD 161832 & 1.19	& --	& 1.55	& 0.95  & AB	 \\
HD 161832$^{({\dagger})}$ & 1.19	& 20	& 1.66	& 0.94  & AB	 \\ 
V* V835 Her$^{({\dagger})}$ & 1.02	& --	& 3.72	& --	& SG	\\
HD 179094 & 1.03	& 10	& 1.60	& 1.10	& BB \\
HD 179094$^{({\dagger})}$ & 1.03	& 10	& 1.82	& 1.12	& BB\\
*f Aql$^{({\dagger})}$ & 1.04	& --	& 1.95	& 0.42$^{({\ast})}$	& BB	 \\
V* V1971 Cyg & 0.99	& --	& 1.91	& 1.36	& BB 	 \\
V* V1971 Cyg$^{({\dagger})}$ & 0.99	 & --	& 2.00	& 1.49	& BB \\
V* V2075 Cyg & 1.00	 & --	& --	& 1.26	& BB 	 \\
V* V2075 Cyg$^{({\dagger})}$ & 1.00	& --	& 2.29	& 1.27	& BB 	  \\
V* IM Peg$^{({\dagger})}$ & 1.06	& --	& 2.40	& 1.60	& BB\\
*lam And & 1.02	& --	& 2.00	& --	& BB	 \\
*lam And$^{({\dagger})}$ & 1.02	& --	& 2.09	& --	& BB	  \\
*alf Sge & 2.95	& --	& 0.51	&0.95$^{(\ast)}$ & RC  \\
*alf Sge$^{({\dagger})}$ & 2.95  & --	 & 0.44	 &0.95$^{(\ast)}$	& RC   \\ 
*eps UMi & 3.50	& --	& 0.90	& --	& BB 	  \\ 
    \bottomrule
  \end{tabular}
  \tablefoot{$^{(\dagger)}$Results obtained from observations with $R\sim68\,000$. Other results were obtained with $R\sim36\,000$. $^{(\ast)}$Upper lithium abundance value. BB and AB: stars below and above the RGB luminosity bump, respectively. RC: clump star. SG: subgiant star. $^{({\ddagger})}$The star is attributed to the thick disc of the Galaxy. }
\end{table}

\begin{table}
\centering
\caption{ NLTE lithium abundances alongside results from \citetalias{Bale2025}.}
\label{tab:cccn2}
\begin{tabular}{lccclc}
\hline
\hline
Name        & Mass  & $^{12}$C/$^{13}$C  & C/N  &  $A(\mathrm{Li})$ & Evol.\\
        &  $M_{\odot}$ &     &      &  NLTE  & stage   \\
\hline
*39 Cet         & 1.10  &--  & 2.19 & 0.49$^{({\ast})}$ & BB \\ 
*39 Cet$^{({\dagger})}$   & 1.00  & --   & 2.51 & 0.43$^{({\ast})}$ & BB \\ 
V* BF Psc$^{({\dagger})}$  & 1.16  &--  & 2.45 & 0.50 & BB  \\ 
V* CL Cam     & 1.20   &  --  & 0.95 & 0.46$^{({\ast})}$  & BB\\ 
V* V403 Aur    & 1.16  & --  & 0.95 & 1.42 & AB \\ 
*11 LMi        & 0.86  & -- & 3.40  & 0.74 & SG\\ 
*eps Leo        & 3.63  & -- & 0.93 & 0.94 & BB\\  
*37 Com        & 1.38  & 7 & 1.60  & 1.42 & RC\\ 
HD 145742      & 1.42  & 12  & 1.02 & 0.38$^{({\ast})}$ & AB \\ 
HD 145742 $^{({\dagger})}$     &  1.46 & 12  & 0.93 & 0.36$^{({\ast})}$ & AB   \\
*29 Dra         & 1.11   & 15  & 1.86 & 0.50 & BB \\ 
*29 Dra$^{({\dagger})}$       & 1.13   & --    & 1.70  & 0.55 & BB \\ 
*62 Ser        & 1.16  & 22  & 1.55 & 1.33 & BB \\ 
*62 Ser$^{({\dagger})}$      & 1.19  & 22  & 1.54 & 1.38  & BB \\  
*b01 Cyg       & 1.19  &--  & 1.47 & 0.70 & BB \\ 
*b01 Cyg$^{({\dagger})}$     & 1.16  & 16  & 1.70 &0.65 & BB \\ 
HD 200740     & 2.17  & 20  & 0.78 & 1.60 & BB \\ 
HD 200740$^{({\dagger})}$   & 2.14  &--  & 0.91 & 1.63 & BB \\ 
V* KX Peg     & 1.06  & 27  & 2.24 & 1.46 & BB \\ 
V* KX Peg$^{({\dagger})}$   & 1.06  & -- & 2.29 & 1.51 & BB \\
\hline
\end{tabular}
\tablefoot{$^{({\dagger})}$Results were obtained from observations with $R\sim68\,000$; other results were obtained with $R\sim36\,000$. $^{(\ast)}$Upper lithium abundance value. BB or AB: stars below or above the RGB luminosity bump, respectively. RC: clump star. SG: subgiant star. The values of mass, $^{12}$C/$^{13}$C, C/N, and the evolutionary stage were taken from \citetalias{Bale2025} for convenience. }
\end{table}

Carbon isotope and C/N ratios are sensitive probes of nuclear burning and mixing inside stars and are often used to trace evolutionary abundance changes. Fig.~\ref{CC} compares the $^{12}$C/$^{13}$C and C/N ratios of the investigated RS\,CVn stars with several recent theoretical predictions. These include the 1DUP and thermohaline-induced extra-mixing (TH) models from \cite{Lagarde2017}, as well as the models that incorporate both thermohaline- and rotation-induced mixing (TH+ROT) in addition to the 1DUP, as presented by \cite{Charbonnel2017}, all for solar-metallicity stars.

In Fig.~\ref{CC}, stars located above the RGB luminosity bump (red circles) and clump stars (red triangles) show evidence of extra-mixing, as expected. However, one star (V*\,OP\,And) belongs to the post-RGB luminosity bump stars, but has a relatively large C/N ratio of 2.60, the highest red circle in the C/N versus mass plot. Its carbon isotope ratio is not greatly reduced, with a value of 24. This K1\,III spectral-type giant also shows a very high lithium abundance and is classified as a Li-rich star in this paper and , for example, in. \cite{Balachandran2000} and \cite{Goncalves2020}. Since the first Li-rich K giant star was discovered by \cite{Wallerstein1982}, a complete explanation of this phenomenon has remained unknown. Thermohaline- or rotation-induced mixing, which predicts lithium production via the Cameron-Fowler mechanism \citep{Cameron&Fowler1971}, is unlikely to apply to V*\,OP\,And, since it should also bring CN-cycled material to the surface, lowering the C/N and \textsuperscript{12}C/\textsuperscript{13}C ratios. Lithium enrichment via mass transfer from a companion is unlikely, as no observations indicate that V*\,OP\,And is a binary. Another explanation for lithium enhancement, such as planet engulfment, could be tested by investigating whether all lithium isotopes are elevated. Therefore, the origin of  V*\,OP\,And as an Li-rich star remains to be investigated.    

\begin{figure*}
    \includegraphics[width=1.0\textwidth]{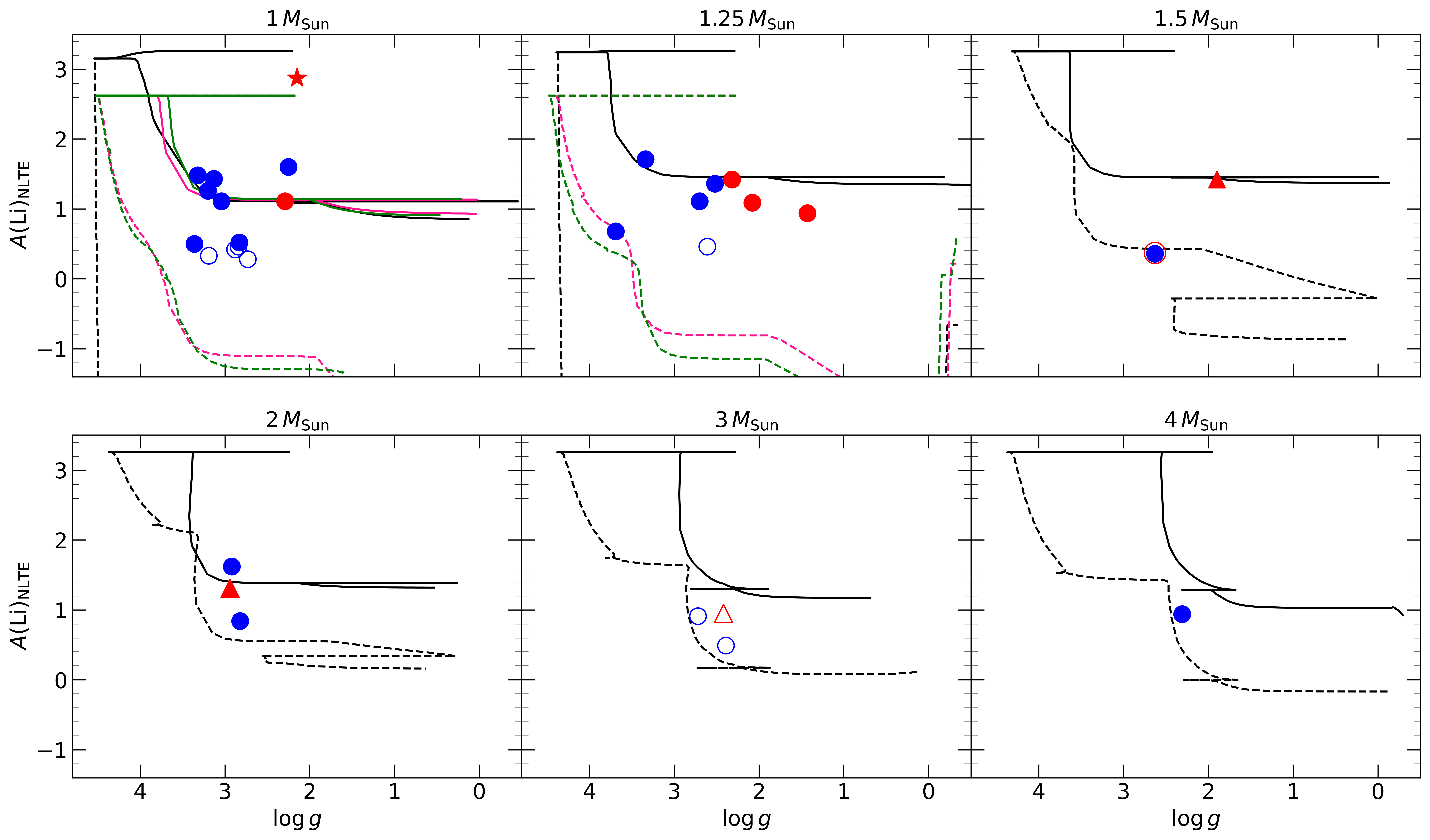}
    \caption{NLTE lithium abundance as a function of log\,$g$ for stars with different masses. The lines represent models for stars evolving from the subgiant branch to the RGB tip; solid lines show the 1DUP, the dashed lines represent the models with thermohaline- and rotation-induced mixing from \cite{Lagarde2012}. Black lines correspond to Z=0.014, pink lines to Z=0.004, and green lines to Z=0.002. Symbol meanings are the same as in Fig.~\ref{cnomg}. Unfilled symbols represent upper lithium abundance limits. The Li-rich star is marked with a red star symbol.}
    \label{limodel}
\end{figure*}

Stars located below the RGB luminosity bump (blue circles) follow the predictions of the 1DUP model, as expected. However, *iot\,Gem and HD\,179094, both situated below the RGB luminosity bump (see Fig.\ref{sett}), exhibit lower \textsuperscript{12}C/\textsuperscript{13}C ratios than predicted by the 1DUP models (cf. \citealt{Lagarde2017}; \citealt{Charbonnel2017}). These stars show carbon isotope ratios typical of stars that have already undergone extra-mixing, which generally begins at the RGB luminosity bump. Specifically, *iot\,Gem and HD\,179094 have \textsuperscript{12}C/\textsuperscript{13}C values of approximately 14 and 10, respectively, consistent with the thermohaline-induced mixing model. Rotation-induced mixing has little influence on stars at these masses. As in \citetalias{Bale2025}, we searched our sample stars for emission in the Ca\,\textsc{ii} lines at 8498, 8542, and 8662~\AA. Emission is detected in HD\,179094. Information on emissions in the Ca\,\textsc{ii} lines in the spectra of other sample stars is provided in Table~\ref{table:Results}.

In our sample of RS\,CVn stars, we identified two additional stars (*iot\,Gem and HD\,179094) that are below the RGB luminosity bump and with  \textsuperscript{12}C/\textsuperscript{13}C ratios significantly lower than the 1DUP values. This increases the number of identified RS\,CVn stars identified as exhibiting extra-mixing that begins earlier than in normal giants. Previously identified examples include *29\,Dra, *b01\,Cyg, V*\,V834\,Her \citepalias{Bale2025}, and $\lambda$\,And \citep{Tautvaisiene2010a, Drake2011}. The star *29\,Dra was also investigated by \cite{Barisevicius2010}. Thus, \textsuperscript{12}C/\textsuperscript{13}C ratios in about 10\% of the investigated RS\,CVn stars indicate the onset of extra-mixing before the RGB luminosity bump. Finally, the C/N ratios in the investigated subgiant stars are higher, as expected for stars before the 1DUP  \citep{Grevesse2007, Asplund2009}.

\subsection{Lithium abundances }
\label{Li-result} 

Lithium abundances are more sensitive to stellar evolution than those of carbon and nitrogen. 
Low- and intermediate-mass giants, having extended convective envelopes, are of particular interest in studies of lithium abundance evolution.  Numerous observational and theoretical studies have been conducted. Most recent large spectroscopic surveys have investigated lithium abundances in large stellar samples (e.g. GALAH, \citealt{Deepak2020, Wang2024}), $Gaia$-ESO (\citealt{Magrini2021, Franciosini2022}), and
LAMOST (\citealt{Zhou2022, Ding2024}).

The peak of RS\,CVn star investigations occurred about 30 years ago, when several groups attempted to determine whether there is any relationship between lithium abundance and stellar magnetic activity (\citealt{Pallavicini1990, Pallavicini1992, Fekel&Balachandra1993, Liu1993, Randich1993, Randich1994}). An overview of lithium abundance investigations from that period is provided by \cite{Pallavicini1994}. At present, with large samples of RGB stars available for comparison, NLTE lithium abundance determinations, and more sophisticated theoretical models of lithium abundance evolution, more comprehensive studies are possible.       

\cite{Lagarde2012} presented a grid of stellar evolutionary models for four metallicities ({\it Z} = 0.0001, 0.002, 0.004, and 0.014) in the mass range 0.85-6.0~$M_\odot$. The models were computed with standard prescriptions or with both thermohaline convection and rotation-induced mixing acting together. The initial rotation velocity of these models on the zero-age main sequence (ZAMS) was set at 45\% of the critical velocity at the corresponding point. The ratio of $V_{\rm ini}/V_{\rm crit} = 0.45$ agrees closely with the mean value of the observed velocity distribution for low- and intermediate-mass stars in young open clusters. This initial rotation rate leads to mean velocities in the main sequence of between 90 and 137~km\,s$^{-1}$. 

Fig.~\ref{limodel} plots the NLTE lithium abundance against surface gravity ($\log g$) for our full sample of stars (Tables~\ref{tab:cccn1} and \ref{tab:cccn2}), together with models from \cite{Lagarde2012} at three metallicities ($Z$ = 0.014, $Z$ = 0.004, $Z$ = 0.002). We divided the stars in our sample by mass and plotted models with several metallicities where necessary. 
About half of the low-mass stars in our sample, for which the lithium abundances were determined, fall close to the predicted post–1DUP plateau. However, other stars, including those where extra-mixing appears to begin earlier, have lower lithium abundances. The intermediate-mass stars have lithium abundances reduced by rotation-induced mixing. In addition, the lithium lines were too weak for reliable abundance determinations in 18 stars. 

\begin{figure}
    \includegraphics[width=\columnwidth]{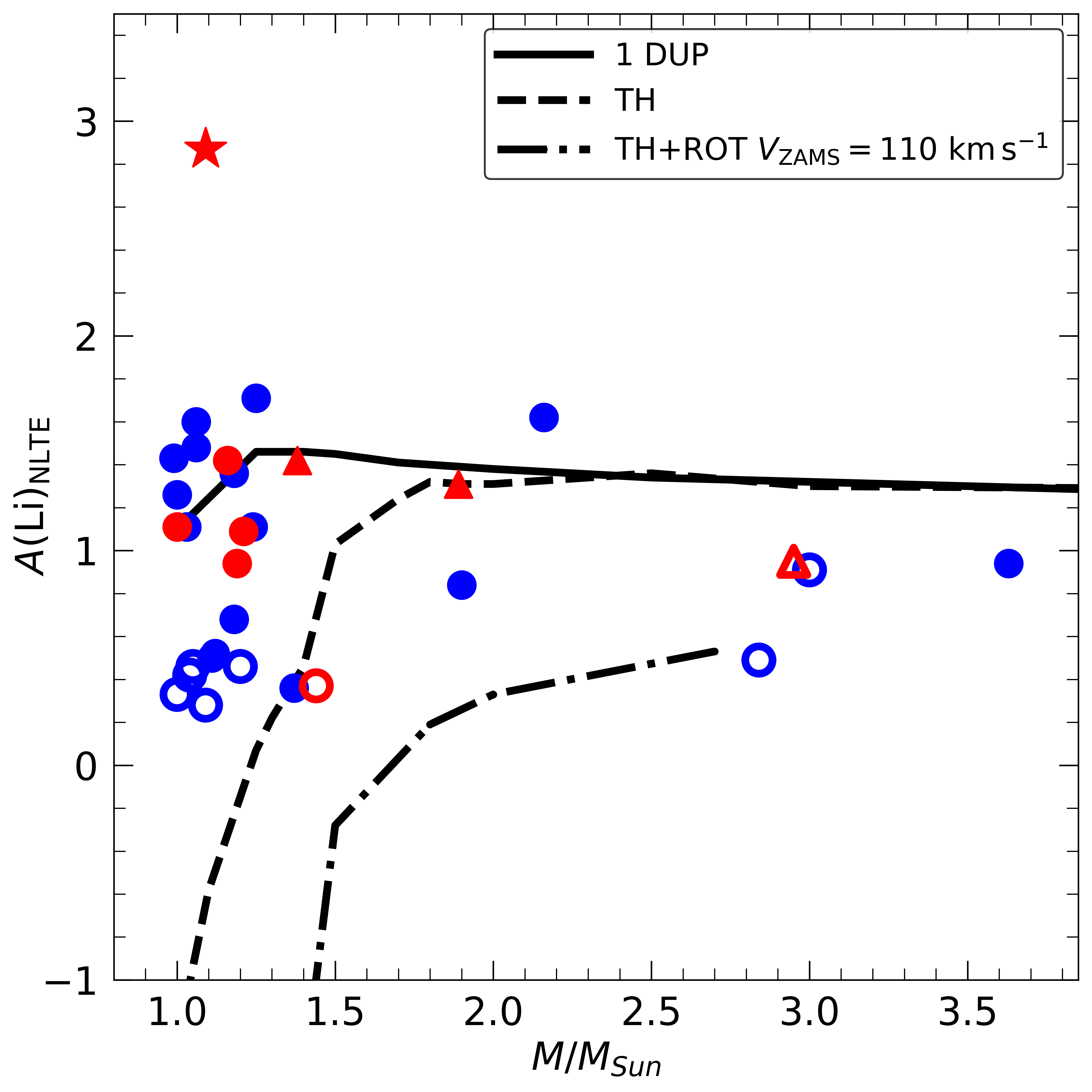}
    \caption{NLTE lithium abundance versus mass.  Symbols are the same as in Fig.~\ref{limodel}. Unfilled symbols denote upper limits. Theoretical models are from \cite{Charbonnel&Lagarde2010}, computed for solar metallicity and mass stars. The solid black line represents lithium abundances  predicted at the 1DUP (1\,DUP), the dashed line corresponds to thermohaline-induced mixing (TH), and the dot-dashed line indicates minimal values in the model with thermohaline- and rotation-induced extra-mixing (TH+ROT) with the rotation on Zero Age Main Sequence (ZAMS) equal to approximately 110~km\,s$^{-1}$.}
    \label{char10}
\end{figure}

\begin{figure}
    \includegraphics[width=\columnwidth]{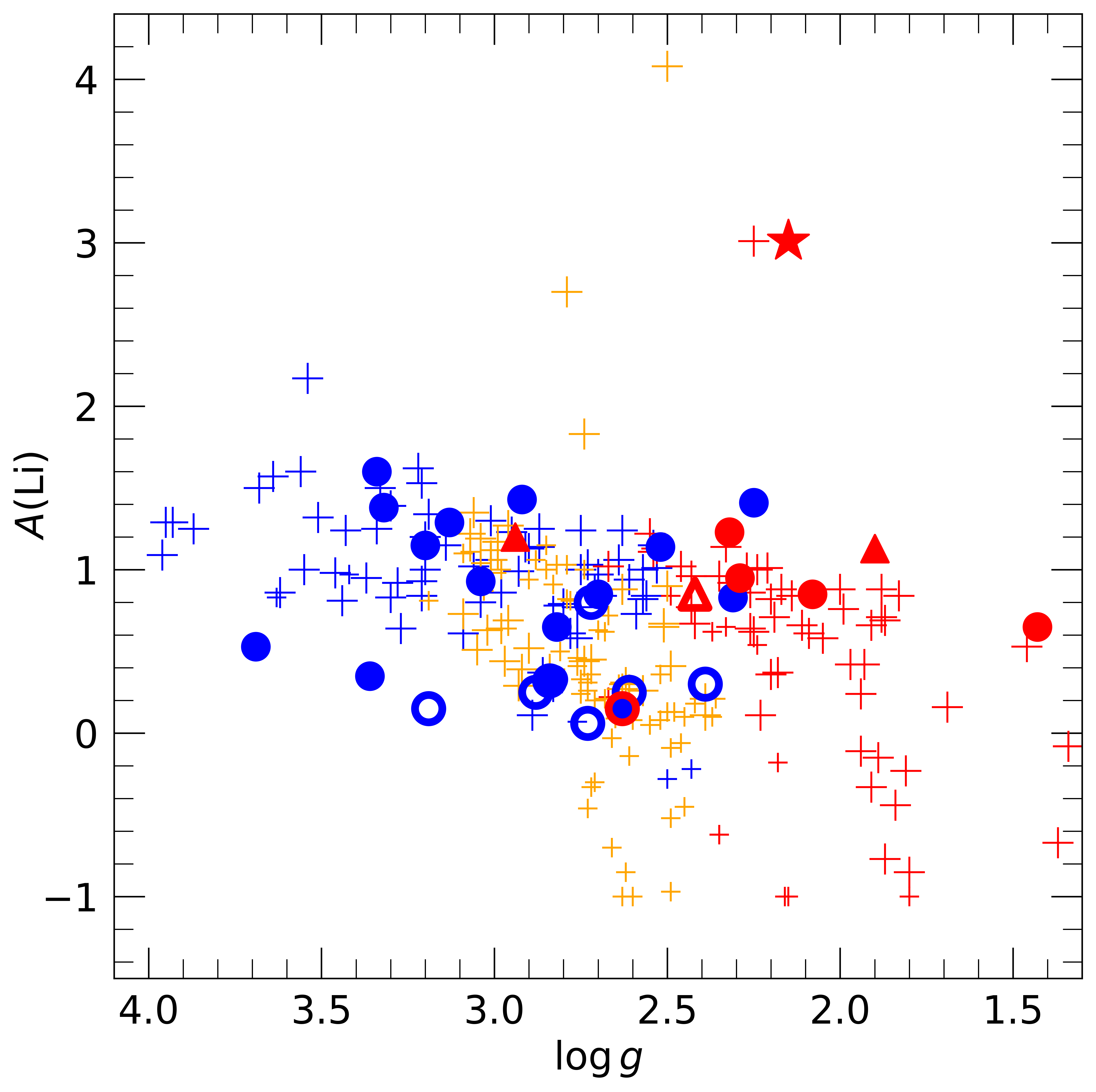}
    \caption{Comparison of LTE lithium abundances as a function of log\,$g$ with results from \cite{Magrini2021}. Symbols are the same as in Fig.~\ref{limodel}. Plus signs indicate  the LTE lithium abundances from \citep{Magrini2021}. The smaller plus signs denote upper limits. Blue symbols mark lower RGB stars, red symbols  indicate upper RGB stars, and yellow symbols represents clump stars. }
    \label{magi21}
\end{figure}

\begin{figure}
    \includegraphics[width=\columnwidth]{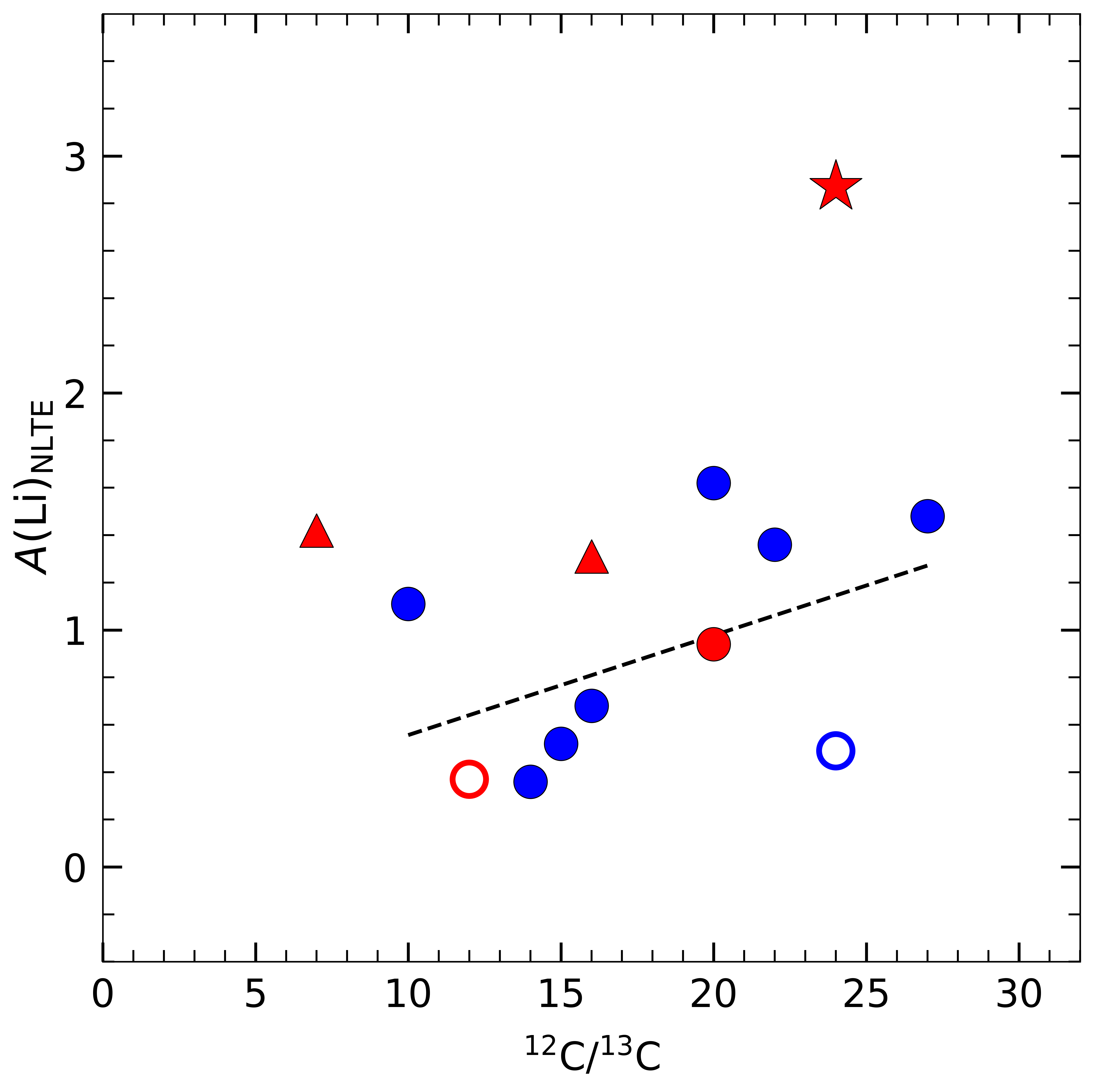}
    \caption{NLTE lithium abundance as a function of \textsuperscript{12}C/\textsuperscript{13}C. Symbols are the same as in Fig.~\ref{limodel}. The dashed line represents a linear, statistically insignificant fit to first-ascent giants, with a PCC of 0.48. }
    \label{ccli}
\end{figure}

Because \cite{Lagarde2012} did not provide a model with purely thermohaline-induced extra-mixing, we compared our data with the models of \cite{Charbonnel&Lagarde2010}, who studied the effects of thermohaline instability and rotation-induced mixing in the 1--4~$M_\odot$ mass range at solar metallicity. Fig.~\ref{char10} compares the NLTE lithium abundances of our sample stars with these models. This comparison shows that extra-mixing in low-mass stars does not exceed pure thermohaline-induced mixing, and confirms that the extra-mixing of lithium in intermediate-mass stars is driven by rotation-induced mixing.

In Fig.~\ref{magi21}, we also compare the lithium abundances of our sample stars with the data from \cite{Magrini2021} . We selected this study because, as in our work,  the investigated giants were divided according to their evolutionary stage (lower RGB, upper RGB, and red clump stars). Because LTE lithium abundances were presented in their work, we compared them with LTE lithium abundances determined in our study. The comparison shows that the RS\,CVn stars follow the overall decline of lithium with evolution driven by 1DUP and extra-mixing, and that among the lower RGB stars there are RS\,CVn stars with lithium depletion caused by extra-mixing or rotation, depending on their mass. 

The lithium abundances in several investigated RS\,CVn subgiants are also reduced, lying between 2.24 and 0.74~dex. However, these values fall in the same region as normal subgiants with similar atmospheric parameters  (see, e.g. Fig.~15 in \citealt{Franciosini2022}). 

Finally, Fig.~\ref{ccli} compares the determined lithium abundances with carbon isotope ratios to test for the presence of a relation. The sample of stars with both $A$(Li) and \textsuperscript{12}C/\textsuperscript{13}C ratios determined is rather small. The dashed line represents a linear, statistically insignificant positive relation for the first-ascent giants, with a Pearson's correlation coefficient ${\rm PCC}=0.48$ and a probability value $p=0.16$. However, this plot highlights the red clump star *37\,Com, which has a low \textsuperscript{12}C/\textsuperscript{13}C of 7 and a relatively high $A$(Li)=1.42. This may indicate that the lithium abundance in this star was enhanced near the RGB tip or during the helium flash. This possibility has recently been widely discussed in the literature (\citealt{Schwab2020, Mori2021, Malick2025, Li2025}, and references therein).

\section{Summary and conclusions} 
\label{conclusion}
In this study, we continue the investigation of chromospherically active RS\,CVn stars.  In \citetalias{Bale2025}, we investigated 20 RS\,CVn stars with an emphasis on determining CNO abundances and carbon isotope ratios. Here, we observe 32 RS\,CVn stars at high spectral resolution with the 1.65~m telescope of the Moletai Astronomical Observatory and the UVES spectrograph to derive their atmospheric parameters and Li, $^{12}$C, $^{13}$C, N, O, and Mg abundances. Lithium abundances were also determined using archival spectra for 13 RS\,CVn stars investigated in \citetalias{Bale2025}. 

From a total sample of 52 investigated RS\,CVn stars, we draw the following main conclusions: 

\begin{itemize}

\item
Among the RS\,CVn giants, some stars exhibit extra-mixing that begins earlier than in normal giants, as indicated by the  lowered \textsuperscript{12}C/\textsuperscript{13}C ratios in *iot\,Gem, HD\,179094, *29\,Dra, *b01\,Cyg, and V*\,V834\,Her, which have not reached the evolutionary phase of RGB  luminosity bump yet. 

\item 
The Li-rich RS\,CVn giant V*\,OP\,And has relatively high C/N and \textsuperscript{12}C/\textsuperscript{13}C ratios, raising questions about the origin of its lithium enhancement. 

\item 
About half of the low-mass giants, for which we determined the lithium abundances, follow the first dredge-up (1DUP) predictions. However, other stars, including those where extra-mixing begins earlier, have lithium abundances reduced by thermohaline-induced mixing. The intermediate-mass stars show lithium abundances reduced by rotation-induced mixing. 

\item 
The comparison with observations of normal giants shows that RS\,CVn stars follow the overall decline of lithium with evolution driven by 1DUP and extra-mixing, and that among the lower RGB stars, some RS\,CVn stars display lithium depletion due to extra-mixing. 
            \end{itemize}

\section*{Data availability}

Full tables~\ref{table:Results} and \ref{table:Results2} are available at the CDS via anonymous ftp.
\begin{acknowledgements}
We acknowledge funding from the Research Council of Lithuania (LMTLT, grant No. S-MIP-23-24). 
This work has made use of data from the European Space Agency (ESA) mission
{\it Gaia} (\url{https://www.cosmos.esa.int/gaia}), processed by the {\it Gaia} Data Processing and Analysis Consortium (DPAC,
\url{https://www.cosmos.esa.int/web/gaia/dpac/consortium}). Funding for the DPAC has been provided by national institutions, in particular, the institutions participating in the {\it Gaia} Multilateral Agreement.
We have made extensive use of the NASA ADS and SIMBAD databases.
\end{acknowledgements}



\bibliographystyle{aa}
\bibliography{ref.bib}

\onecolumn
\begin{appendix}

\section{Machine readable tables of results}

 \begin{longtable}{llll}
 \caption{Determinations of stellar main parameters and C, N, O, and Mg abundances }
 \label{table:Results}\\
 \hline
 \hline
 Col & Label & Units & Explanations\\
 \hline
 1      & ID                 & --          & Tycho-2 catalogue identification\\
 2   & Name               & --           & Stellar name \\
 3   & Res                & --          & Spectral resolution   \\
 4      & $T_{\rm eff}$     & K             & Effective temperature\\
 5      & $e$\_$T_{\rm eff}$  & K        & Uncertainty in effective temperature\\
 6      & Log\,$g$               & dex & Stellar surface gravity\\
 7      & $e$\_Log\,$g$            & dex & Uncertainty in stellar surface gravity\\
 8      & [Fe/H]             & dex          & Metallicity\\
 9      & $e$\_Fe\,\textsc{i}          & dex        & Uncertainty in [Fe\,\textsc{i}/H]\\
 10   & n\_Fe\,\textsc{i}             &  --         & Number of Fe\,\textsc{i} lines \\
 11  & $e$\_Fe\,\textsc{ii}            &  dex        & Uncertainty in [Fe\,\textsc{ii}/H] \\
 12  & n\_Fe\,\textsc{ii}            &   --          & Number of Fe lines \\
 13     & $V_{\rm t}$         &  km\,s$^{-1}$   & Microturbulence velocity\\
 14     & $e\_Vt$              & km\,s$^{-1}$  & Uncertainty in microturbulence velocity\\
 15     & [C/H]                  & dex              & Carbon abundance\\
 16     & $e$\_[C/H]         & dex          & Uncertainty in carbon abundance\\
 17  & n\_C                  & --           & Number of C$_2$ lines \\
 18     & [N/H]                  & dex              & Nitrogen abundance\\
 19     & $e$\_[N/H]         & dex          & Uncertainty in nitrogen abundance\\
 20  & n\_N                  & --           & Number of CN lines \\
 21     & [O/H]                  & dex              & Oxygen abundance\\
 22     & $e$\_[O/H]         & dex          & Uncertainty in oxygen abundance\\
 23  & n\_O                  & --           & Number of oxygen lines \\
 24     & [Mg/H]             & dex          & Magnesium abundance\\
 25     & $e$\_[Mg/H]        & dex          & Uncertainty in magnesium abundance\\
 26  & n\_Mg                  & --           & Number of magnesium lines \\   
 27   & $^{12}$C/$^{13}$C & --           & Carbon isotope ratio \\
 28   & $e$\_$^{12}$C/$^{13}$C & --        & Uncertainty in carbon isotope ratio \\
 29     & C/N               & --           & Carbon-to-nitrogen abundance ratio \\
 30    & Mass              & $M_\odot$    & Stellar mass \\
  31    & $e$\_Mass         & $M_\odot$    & Uncertainty of mass \\
  32    & Evol              & --    & Evolutionary stage (BB or AB -- stars below or above the RGB luminosity bump,\\
        &   &  & respectively. RC -- Clump stars, SG  -- subgiant stars)\\
  33    & Emission          &      --      & Yes -- if emission is visible in the Ca\,\textsc{ii} lines \\
 \hline
 \end{longtable}
 \tablefoot{Full table is available at the CDS}
 \centering

 \begin{longtable}{llll}
 \caption{Determinations of lithium abundances}
 \label{table:Results2}\\
 \hline
 \hline
 Col & Label & Units & Explanations\\
 \hline
 1      & ID                 & --          & Tycho-2 catalogue identification\\
 2   & Name               & --           & Stellar name \\
 3   & Res                & --          & Spectral resolution   \\
 4     & $A$(Li)\_LTE         &  dex & Lithium abundance in LTE \\
 5     & $A$(Li)\_NLTE     & dex  & Lithium abundance in NLTE \\
 6     & $e$\_$A$(Li)    & dex  & Uncertainty in lithium abundance \\
 7  & \textup{Flag\_}\textit{A}\textup{(Li)} &  -- & [0/1] Limit flag on $A$(Li) (1 for upper limit)\\
 \hline
 \end{longtable}
 \tablefoot{Full table is available at the CDS}


\clearpage
\section{Atmospheric parameters of investigated stars}

\begin{center}
\captionof{table}{Main atmospheric parameters}
\label{table:atmospheric_parameters}
\begin{tabular}{lccccrcrcccc}
 \hline
 \hline
Star      & $T_{\rm eff}$ & $e_{T_{\rm eff}}$ & log\,$g$ & $e_{{\rm log}g}$ & [Fe/H] & $e_{\rm Fe\,I}$ & n$_{\rm Fe\,I}$ & $e_{\rm Fe\,II}$ & n$_{\rm Fe\,II}$ & $V_{\rm t}$ & $e_{V_{\rm t}}$ \\ 
    &  K & K &   &  &  &  &  &  &  & km\,s$^{-1}$  &  km\,s$^{-1}$ \\
 \hline 
*33 Psc & 4660 & 55      & 2.73  & 0.15    & $-0.21$ & 0.06  & 64   & 0.03    & 6  & 1.00  & 0.21  \\ 
HD 6497   & 4345	& 50	& 2.37	& 0.16	& $-0.07$	& 0.10 & 72	& 0.08	& 5	& 1.12	& 0.22 \\
HD 6497$^{({\dagger})}$  & 4335	& 45	& 2.31	& 0.14	& $-0.02$	& 0.08	& 74	& 0.07	& 5	& 1.02	& 0.19  \\ 
V* OP And$^{({\dagger})}$ & 4466	& 55	& 2.15	& 0.17	& $-0.40$	& 0.10	& 71	& 0.05	& 6	& 1.77	& 0.18\\ 
V* V1149 Ori & 4613	& 58	& 2.08	& 0.18	& $-0.42$	& 0.10	& 65	& 0.06	& 7	 & 1.52	& 0.18  \\
*iot Gem & 4801	& 50	& 2.63	& 0.15	& $-0.07$	& 0.08	& 64	& 0.07	& 6	 & 1.32	& 0.17 \\ 
* sig Gem$^{({\dagger})}$  & 4620	& 85	& 2.70	& 0.07	& 0.08	& 0.08	& 27	& 0.09	& 7	& 1.64	& 0.26 \\ 
* bet Gem  & 4868	& 45	& 2.82	& 0.15	& 0.04	& 0.08	& 72	& 0.04	& 7	& 1.33	& 0.17\\ 
* g Gem  & 4125	& 80	& 1.64	& 0.16	& $-0.35$	& 0.10	& 58	& 0.03	& 5 	& 1.49	& 0.18\\
HD 71028 & 4580	& 60	& 1.75	& 0.19	& $-0.77$ 	& 0.10	& 60	& 0.10	& 6	 & 1.31	& 0.20\\ 
V* FI Cnc & 5175	& 55	& 3.34	& 0.25	& $-0.11$	& 0.12	& 75	& 0.08	& 9	& 1.43	& 0.24\\ 
* 10 LMi & 5061	& 50	& 2.89	& 0.16	& $-0.08$	& 0.07	& 58	& 0.05	& 6	& 1.31	& 0.20\\
V* FG UMa & 4697 & 48	& 2.29	& 0.18	& $-0.85$	& 0.09	& 55	& 0.08  	& 5	& 1.84	& 0.19\\ 
* tau Leo  & 4953	& 40	& 2.39	& 0.16	& $-0.02$	& 0.08	& 75	& 0.09	& 7	& 1.59	& 0.15\\ 
* 93 Leo$^{({\dagger})}$ & 5130	& 45	& 2.94	& 0.16	& $-0.42$	& 0.07 	& 58	& 0.04	& 6	& 0.48	& 0.33\\
* c Vir  & 4500	& 48	& 2.04	& 0.11	& $-0.44$	& 0.06	& 58	& 0.07  	& 6	 & 1.60	& 0.12  \\ 
V* BM CVn  & 4816	& 85	& 3.20	& 0.07	& 0.15	& 0.08	& 35	& 0.07	& 5	& 1.89	& 0.28  \\ 
V* IT Com  & 4705	& 70	& 3.19	& 0.18	& $-0.30$	& 0.09	& 66	& 0.07	& 6	& 1.55	& 0.23 \\ 
* 7 Boo & 5367	& 50	& 2.82	& 0.22	& $-0.04$	& 0.10	& 50	& 0.09	& 7	& 1.87	& 0.21 \\ 
* 7 Boo$^{({\dagger})}$ & 5313	& 50	& 2.61	& 0.21	& $-0.07$	& 0.09	& 57	& 0.08	& 6	& 1.83	& 0.20 \\ 
* eta Boo  & 6220	& 60	& 3.95	& 0.21	& 0.34	& 0.08	& 71	& 0.07	& 9	& 1.59	& 0.23 \\ 
* eta Boo$^{({\dagger})}$  & 6158	& 55	& 3.96	& 0.20	& 0.25	& 0.08	& 69	& 0.09	& 7	& 1.67	& 0.22  \\ 
V* FR Boo & 4355	& 70	& 2.29	& 0.17	& $-0.35$	& 0.10	& 52	& 0.08  	& 6	& 1.57	& 0.21\\ 
V* HK Boo$^{({\dagger})}$ & 4948	& 50	& 3.41	& 0.19	& $-0.67$	& 0.09	& 61	& 0.08	& 6	& 1.16	& 0.23 \\ 
HD 141690 & 6085	& 88	& 4.27	& 0.12	& $-0.29$	& 0.09	& 42	& 0.06	& 5	& 1.99	& 0.60  \\
HD 161832  & 4420	& 80	& 1.41	& 0.16	& $-0.59$	& 0.08	& 35	& 0.08  	& 5	& 1.77	& 0.16 \\
HD 161832$^{({\dagger})}$ & 4430	& 65	& 1.44	& 0.16	& $-0.59$	& 0.08	& 43	& 0.08	& 4	& 1.81	& 0.16 \\ 
V* V835 Her$^{({\dagger})}$ & 5280	& 60	& 3.86	& 0.21	& $-0.30$	& 0.10	& 60	& 0.10	& 5	& 1.57	& 0.30 \\
HD 179094 & 4758  & 58	& 3.03	& 0.19	& $-0.25$	& 0.10	& 58	& 0.08	& 7	& 1.70	& 0.24 \\
HD 179094$^{({\dagger})}$ & 4760 & 62	& 3.05	& 0.19	& $-0.30$	& 0.10	& 58	& 0.06	& 6	& 1.85	& 0.24 \\
* f Aql$^{({\dagger})}$ & 4827	& 35	& 2.88	& 0.11	& $-0.28$	& 0.05	& 72	& 0.09	& 7	& 0.84	& 0.14 \\
V* V1971 Cyg  & 4885	& 55	& 3.15	& 0.18	& $-0.47$	& 0.09	& 48	& 0.08	& 7	& 1.68	& 0.22 \\
V* V1971 Cyg$^{({\dagger})}$ & 4815	& 72	& 3.10	& 0.19	& $-0.43$	& 0.10	& 46	& 0.10	& 6	& 1.67	& 0.26\\
V* V2075 Cyg & 5040	& 65	& 3.29	& 0.21	& $-0.52$	& 0.10	& 48	& 0.10	& 6	& 1.86	& 0.28 \\
V* V2075 Cyg$^{({\dagger})}$ & 5020	& 52	& 3.10	& 0.19	& $-0.51$	& 0.09	& 49	& 0.10	& 4	& 1.87	& 0.24\\
V* IM Peg$^{({\dagger})}$ & 4820	& 59	& 2.25	& 0.19	& $-0.26$	& 0.10	& 47	& 0.12	& 8	& 1.89	& 0.19\\
* lam And & 4750	& 35	& 2.73	& 0.17	& $-0.50$	& 0.09	& 80	& 0.08	& 7	& 1.14	& 0.18 \\
* lam And$^{({\dagger})}$ & 4732	& 30	& 2.80	& 0.13	& $-0.56$	& 0.07	& 81	& 0.09	& 7	& 1.25	& 0.13\\
* alf Sge & 5610	& 55	& 2.48	& 0.09	& 0.03	& 0.08	& 64	& 0.05	& 6	& 1.78	& 0.22 \\
* alf Sge$^{({\dagger})}$ & 5630 & 45	& 2.36	& 0.07	& 0.03	& 0.06	& 63	& 0.09	& 6	& 1.83	& 0.19\\ 
* eps UMi & 5279	& 55	& 2.48	& 0.10	& 0.10	& 0.10	& 49	& 0.09	& 6	& 1.47	& 0.20 \\ 
\hline
\end{tabular}
\tablefoot{$^{({\dagger})}$Results obtained from observations with $R\sim68\,000$. Other results were obtained with $R\sim36\,000$.}
\end{center}

\end{appendix}
\end{document}